\documentclass[%
 reprint,
superscriptaddress,
 amsmath,amssymb,
 prl,
floatfix,
]{revtex4-2}

\usepackage{graphicx}
\usepackage{dcolumn}
\usepackage{bm}
\usepackage{booktabs}
\usepackage[dvipsnames]{xcolor}
\usepackage[colorlinks, allcolors=NavyBlue]{hyperref}
\renewcommand{\bf}[1]{\mathbf{#1}}

\newcommand{\mlsc}{\alpha_0} 

\begin{document}

\title{Entropy Production Bounds the Accuracy of Computation in Markov Networks}

\author{Songela W. Chen}
\affiliation{Department of Chemistry, University of California, Berkeley, CA, 94720, USA}
\affiliation{Kavli Energy NanoScience Institute, University of California, Berkeley, CA, 94720, USA}
\author{David T. Limmer}\email{dlimmer@berkeley.edu}
\affiliation{Department of Chemistry, University of California, Berkeley, CA, 94720, USA}
\affiliation{Kavli Energy NanoScience Institute, University of California, Berkeley, CA, 94720, USA}
\affiliation{Chemical Sciences Division, Lawrence Berkeley National Laboratory, Berkeley, CA 94720, USA}
\affiliation{Materials Sciences Division, Lawrence Berkeley National Laboratory, Berkeley, CA 94720, USA}

\date{\today}

\begin{abstract}
Biological and artificial networks compute by transforming time-dependent inputs into functional outputs. Because the internal state of a stochastic network relaxes on finite timescales, its output generally lags behind a changing environment, producing computational errors. We show that for reversible continuous-time Markov networks the error admits a universal thermodynamic bound. Decomposing the total error into representation and lag contributions, we derive an inequality relating the lag error to the entropy production rate and a memory time equal to the integrated equilibrium autocorrelation of the output observable. The bound implies that accurate dynamical computation requires either substantial dissipation or long-lived memory encoded in slowly relaxing modes. We demonstrate these principles in artificial Markov networks and in models of biochemical information processing. Our results establish a thermodynamic limit on information processing in stochastic networks and provide a quantitative framework for understanding the energetic costs of biological computation.
\end{abstract}

\maketitle

Living systems continuously process information from fluctuating environments. Cells infer nutrient concentrations, bacteria collectively estimate population density through quorum sensing, and photosynthetic organisms regulate photoprotection in response to changing illumination \cite{tkavcik2016information,short2022xanthophyll,short2023kinetics,feng2015qrr,waters2005quorum,govern2014optimal,tkavcik2025information}. In each case, a biochemical network performs a computation, transforming a time-dependent input into a functional output \cite{phillips2012physical}. Similar input-output transformations are increasingly engineered in synthetic chemical circuits and neuromorphic devices, where stochastic reaction networks are trained to emulate prescribed dynamical behaviors \cite{arcadia2021leveraging,floyd2025limits}.
The ability of a stochastic network to compute accurately is fundamentally constrained by its internal dynamics. Because information propagates through a network on finite timescales, its state generally lags behind a changing environment \cite{crooks2007measuring,sivak2012thermodynamic}. This lag leads to errors in the output and necessitates the continual dissipation of free energy. 
Thermodynamic constraints on nonequilibrium processes have been extensively studied through fluctuation relations, thermodynamic uncertainty relations, and speed limits, which respectively bound the probabilities of rare events, the precision of currents, and the speed of state transformations \cite{limmer2024statistical,peliti2021stochastic,horowitz2020thermodynamic}. These results reveal deep connections between dissipation and dynamical performance \cite{horowitz2010nonequilibrium,shiraishi2016universal,shiraishi2018speed,barato2016cost,still2012thermodynamics,helms2025stochastic,kuznets2021dissipation}. However, they do not  address the question of how accurately a stochastic network can realize a desired input-output transformation.

In this Letter, we derive a thermodynamic bound on the computational error of reversible Markov networks driven by time-dependent signals. We show that the total error naturally separates into a representation error, determined by the ability of the instantaneous steady state to encode the desired input-output map, and a lag error arising from the finite relaxation of the probability distribution. We then demonstrate that the lag error is generally bounded by the entropy production rate, and takes a simple form near equilibrium 
\begin{equation}
\varepsilon_{\mathrm{lag}}^2 \le \tau_\mathrm{m} \sigma ,
\label{eq:main_bound}
\end{equation}
where $\sigma$ is the entropy production rate and $\tau_\mathrm{m}$
is a time integrated susceptibility determined entirely by equilibrium fluctuations.
This bound establishes a thermodynamic limit on computation in stochastic networks. The susceptibility quantifies the extent to which information on the output observable is stored in slowly relaxing modes of the network and therefore plays the role of a computational memory. Accurate computation consequently requires either dissipation or long-lived memory. Unlike earlier learning-rate bounds, which quantify the thermodynamic cost of reducing uncertainty about an external stochastic process~\cite{barato2014efficiency}, Eq.~\ref{eq:main_bound} constrains the lag error of a specified output observable, with $\tau_\mathrm{m}$ selecting the relaxation modes relevant to the computation.We demonstrate these principles in trained Markov networks and in models of biochemical information processing including ligand sensing \cite{einav2017monod}, gene regulation \cite{razo2018tuning}, quorum sensing \cite{mehta2009information}, and nonphotochemical quenching \cite{short2022xanthophyll}.

We consider a continuous-time Markov process with probabilities
$\mathbf{p}(t)=\{p_1(t),\dots,p_N(t)\}$ that evolves as 
\begin{equation}
\dot{\mathbf{p}}(t)=\mathbf{p}(t)\mathbf{W}[\lambda(t)],
\label{eq:master}
\end{equation}
where $\lambda(t)$ is a time-dependent input signal and
$\mathbf{W}[\lambda(t)]$ is the corresponding generator. For fixed input, the dynamics satisfy detailed balance,
\begin{equation}
\boldsymbol{\pi}(\lambda)\mathbf{W}[\lambda]=0,
\qquad
\pi_i(\lambda)k_{ij}[\lambda]=\pi_j(\lambda)k_{ji}[\lambda],
\label{eq:db}
\end{equation}
with equilibrium distribution $\boldsymbol{\pi}(\lambda)$ and transition rate from $i$ to $j$, $k_{ij}[\lambda]=(\mathbf{W})_{ij}$.
The network is tasked with implementing a map $\mathcal{G}$ from $\lambda(t)$ to an output $\hat r(t)$, such as $\hat r(t)=\mathcal G[\lambda(t)]$,
which may depend on the history of the input. We will assume the network output is read from a subset of states. The instantaneous computational error is then
\begin{equation}
\varepsilon (t)=   r(t)-\hat r(t)\, , \qquad \hat r(t)=\mathbf{a}\cdot \mathbf{p}(t)
\label{eq:error}
\end{equation}
where $r(t)$ and $\hat r(t)$ indicate the target and instantaneous output signal, and $a_i\in{0,1}$ indicates the output states.

We will consider a decomposition of the error into two sources. For a fixed network we define a representation error by comparing the instantaneous steady-state at fixed $\lambda$ to the desired output,
\begin{equation}
\varepsilon_\mathrm{rep}(t) =r(t) -r_\mathrm{eq}[\lambda(t)]\,, \qquad r_\mathrm{eq}[\lambda(t)] = \mathbf{a}\cdot \boldsymbol{\pi}[\lambda(t)]
\end{equation}
which quantifies the inability of the instantaneous equilibrium state to realize the desired input-output relation. A second source of error is that associated with the lag of the evolved probability distribution with its instantaneous steady-state. This lag error is quantified by
\begin{equation}
\varepsilon_\mathrm{lag}(t) = \mathbf{a}\cdot \left\{\bf{p}(t)-\boldsymbol{\pi}[\lambda(t)] \right \}
\end{equation}
such that the total error is $\varepsilon=\varepsilon_\mathrm{rep}+\varepsilon_\mathrm{lag}$ and
\begin{equation}
\varepsilon^2(t) \leq 2\varepsilon_\mathrm{lag}^2(t)+2\varepsilon_\mathrm{rep}^2(t)
\end{equation}
the mean squared error is bounded by a sum of twice the mean squared lag error and representation error.

While the representation error is fundamentally determined by the topology of the network and the model of the dependence of the rates on the input, the lag error has a thermodynamic interpretation. 
Since the observable vector satisfies $a_i\in{0,1}$,
\begin{equation}
|\varepsilon_{\mathrm{lag}}| \le \frac12 \sum_i |p_i-\pi_i|.
\end{equation}
Pinsker's inequality \cite{csiszar2011information} therefore gives
\begin{equation}
\varepsilon_{\mathrm{lag}}^2 \le \frac12 D[\mathbf{p}||\boldsymbol{\pi}],
\label{eq:pinsker}
\end{equation}
where $D[\mathbf{p}||\boldsymbol{\pi}]= \sum_i p_i\ln p_i/\pi_i$ is the relative entropy between the instantaneous distribution and the frozen equilibrium state. For reversible Markov dynamics at a fixed input, the relative entropy decays according to \cite{seifert2012stochastic}
\begin{equation}
\frac{d}{dt}D[\mathbf{p}||\boldsymbol{\pi}] =-\sigma,
\end{equation}
where for the instantaneous frozen-generator 
\begin{equation}
\sigma =\frac12 \sum_{ij} \pi_i k_{ij} (f_i-f_j) \ln\frac{f_i}{f_j},
\label{eq:full_sigma}
\end{equation}
is the entropy production rate and $f_i=p_i/\pi_i$. Combining Eq.~\ref{eq:pinsker} with the modified log-Sobolev inequality for reversible Markov chains~\cite{diaconis1996logarithmic} $D[\mathbf{p}||\boldsymbol{\pi}]\le \sigma/2\alpha$ 
provides
\begin{equation}
\varepsilon_{\mathrm{lag}}^2\le \frac{\sigma}{4\mlsc},
\label{eq:sobolev_bound}
\end{equation}
where $\mlsc$ is the log-Sobolev constant, details of which are discussed in the Supplemental Materials (SM) \cite{SM}.

While Eq.~\ref{eq:sobolev_bound} is general, relating the lag error to the entropy production, it is  difficult to use in practice because the constant $\mlsc$ is not known explicitly \footnote{The constant $\mlsc$ can be estimated from the gap in the generator, $\omega$ and the minimum over $\pi$, $\mlsc \approx \omega \, \mathrm{min}_i (\pi_i)/2$.}. A more physically transparent bound emerges in the regime where the lag from equilibrium is small. To this end, we define
\begin{equation}
h_i= \frac{p_i-\pi_i}{\pi_i},
\qquad
p_i=\pi_i(1+h_i),
\end{equation}
and introduce the centered observable
$\bar a_i= a_i-\mathbf{a} \cdot \boldsymbol{\pi}$.
Next, we define a committor-like function $\mathbf{q}$ \cite{vanden2010transition} through the Poisson equation
\begin{equation}
\mathbf{W} \mathbf{q}=-\bar{\mathbf{a}},
\qquad
\boldsymbol{\pi} \cdot \mathbf{q}=0.
\label{eq:poisson}
\end{equation}
subject to the orthogonality of $\boldsymbol{\pi}$. 
Then the lag error can be written in terms of the same edge-wise differences that control relaxation,
\begin{equation}
\varepsilon_{\mathrm{lag}}=\frac12\sum_{ij}\pi_i k_{ij}(q_i-q_j)(h_i-h_j) \, .
\end{equation}
Applying the Cauchy-Schwarz inequality yields
\begin{equation}
\varepsilon_{\mathrm{lag}}^2 \le \left[\frac12 \sum_{ij} \pi_i k_{ij} (q_j-q_i)^2 \right]
\left[\frac12\sum_{ij} \pi_i k_{ij} (h_i-h_j)^2 \right]. \nonumber
\label{eq:cs}
\end{equation}
The two terms in the inequality each have a simple interpretation. The second term is an approximation to the entropy production,
\begin{equation}
\sigma = \frac12 \sum_{ij} \pi_i k_{ij}(h_i-h_j)^2 + O(h^3),
\label{eq:quadratic_sigma}
\end{equation}
for small deviations from equilibrium. The first term can be rewritten using Eq.~\ref{eq:poisson},  
\begin{equation}
\tau_\mathrm{m}=\sum_i \pi_i \bar a_i q_i = \int_0^\infty dt \, \langle \bar{\mathbf{a}}(0)\cdot \bar{\mathbf{a}}(t)\rangle
\end{equation}
since
\begin{equation}
\mathbf{q} =-\mathbf{W}^{-1}\bar{\mathbf{a}} = \int_0^\infty dt e^{\mathbf{W}t}\bar{\mathbf{a}},
\end{equation}
which we can identify as an observable susceptibility \cite{kipnis1986central}.
We therefore obtain, to leading order in the lag, $\varepsilon_{\mathrm{lag}}^2 \le \tau_\mathrm{m} \sigma \, ,$ which constitutes the central result of this work.
The constant $\tau_\mathrm{m}$ is the integrated equilibrium autocorrelation of the output observable \cite{govern2014optimal,sartori2015free}. It quantifies how strongly the network stores information about the observable in slowly relaxing modes and can be interpreted as a computational memory capacity. Equation \ref{eq:main_bound} thus implies that accurate dynamical computation requires either significant dissipation or long-lived memory.

\begin{figure}
    \centering
    \includegraphics[width=1.\linewidth]{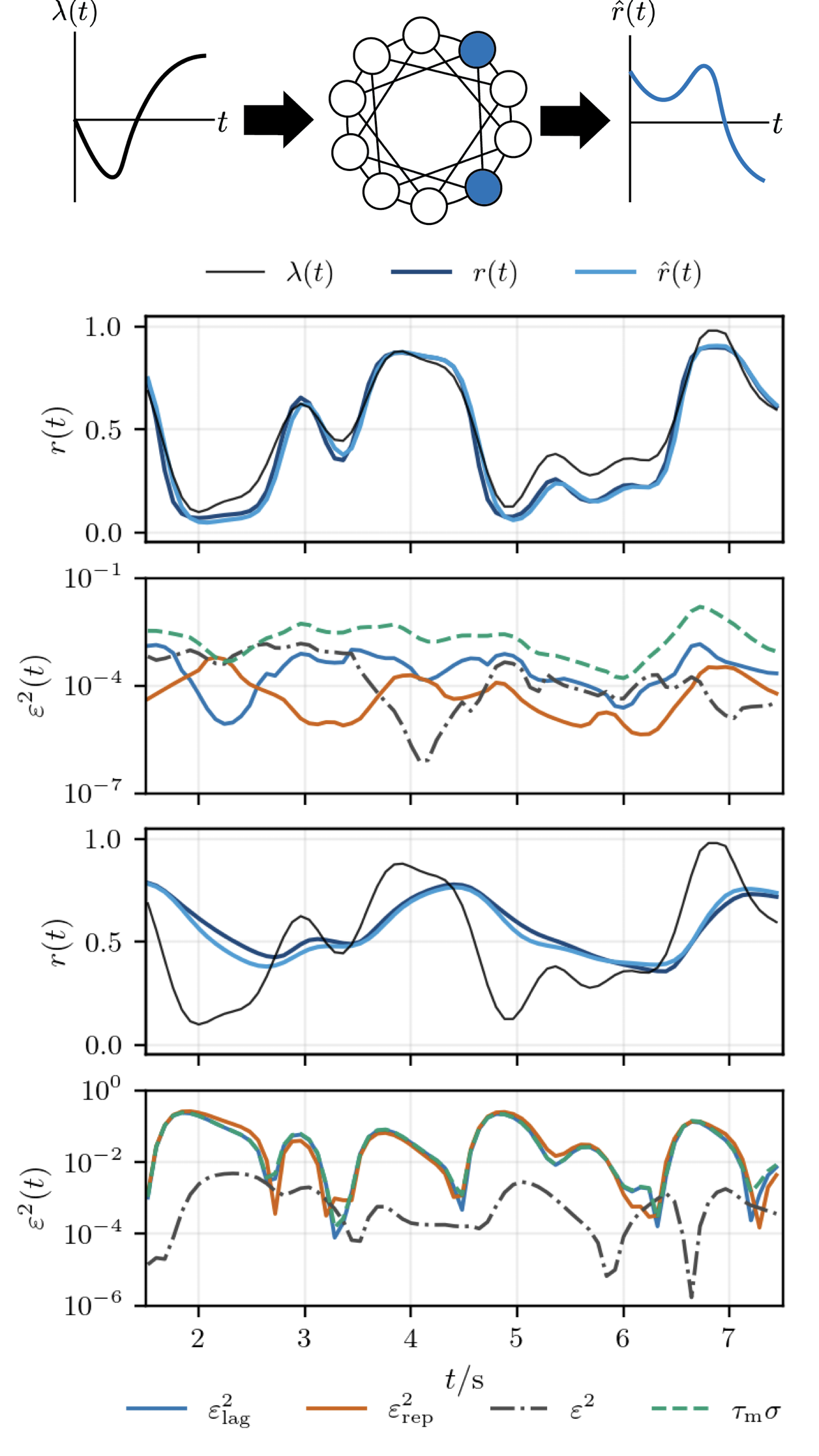}
\caption{
Error decomposition for trained reversible Markov networks with $N=10$ states. The upper pair of panels shows a memoryless target. The lower pair shows a history-dependent target. In each trajectory panel, black denotes the input, dark blue the target output, and light blue the trained network output. The error panels decompose the squared tracking error into lag error $\varepsilon_{\mathrm{lag}}^2$, representation error $\varepsilon_{\mathrm{rep}}^2$, total error $\varepsilon^2$, and the thermodynamic bound $\tau_\mathrm{m}\sigma$. 
}
\label{fig:trained_decomposition}
\end{figure}

To illustrate the consequences of Eq.~\ref{eq:main_bound}, we first consider reversible continuous-time Markov networks trained to approximate prescribed input-output transformations \cite{bengio1996input}. The networks consist of $N$ states arranged on a sparse graph with nearest- and third-nearest-neighbor connections. The transition rates satisfy detailed balance with respect to an input-dependent energy landscape,
\begin{equation}
k_{ij}(\lambda)
=\exp\left \{
g_{ij}(\lambda)
+\frac{\beta}{2}
\left [E_i(\lambda)-E_j(\lambda) \right ]
\right \},
\end{equation}
where $\beta=k_\mathrm{B} T$ is the inverse temperature times Boltzmann's constant. The energies $E_i$ and barrier heights $g_{ij}$ are optimized so that the network output tracks a prescribed signal.

We consider two classes of computation. The first is a memoryless target,
\begin{equation}
r(t)=f[\lambda(t)],
\end{equation}
for which the desired output is a single-valued function of the instantaneous input. The second is a history-dependent target,
\begin{equation}
\dot r = \frac{1}{\tau(t)} \left \{ f[\lambda(t)]-r \right \},
\end{equation}
which depends explicitly on the past trajectory of the input and therefore requires dynamical memory.
Figure~\ref{fig:trained_decomposition} compares the trained responses for the two tasks. For the memoryless target, the equilibrium mapping closely reproduces the desired input-output relation. The overall error is comparable to the representation error, and in this regime, the lag error is small.

The history-dependent task exhibits qualitatively different behavior. Since the target is not a function of the instantaneous input alone, the equilibrium response cannot represent the desired transformation. The network therefore exploits its own relaxation dynamics as a computational resource. The representation and lag errors are individually large but strongly anticorrelated, leading to a total error that is substantially smaller than either contribution alone, $\epsilon^2 \ll \epsilon_{\mathrm{rep}}^2 + \epsilon_{\mathrm{lag}}^2$. In this regime, lag constitutes the memory required to implement the desired computation. Figure~\ref{fig:trained_decomposition} also shows the lag error plotted against the  thermodynamic bound of Eq.~\ref{eq:main_bound} for the trained network. In both tasks, the quadratic approximation         to the entropy is quantitatively consistent with the full entropy calculation. The lag error remains below the corresponding time-dependent bound, with the inequality becoming tight when the nonequilibrium displacement $(\mathbf{p}(t)-\boldsymbol{\pi}[\lambda(t)])$ aligns with the output-relevant relaxation mode encoded by $\mathbf{q}[\lambda(t)]$. In this limit, the dissipation contributes to reducing computational error.

\begin{figure*}
    \centering
    \includegraphics[width=1.\linewidth]{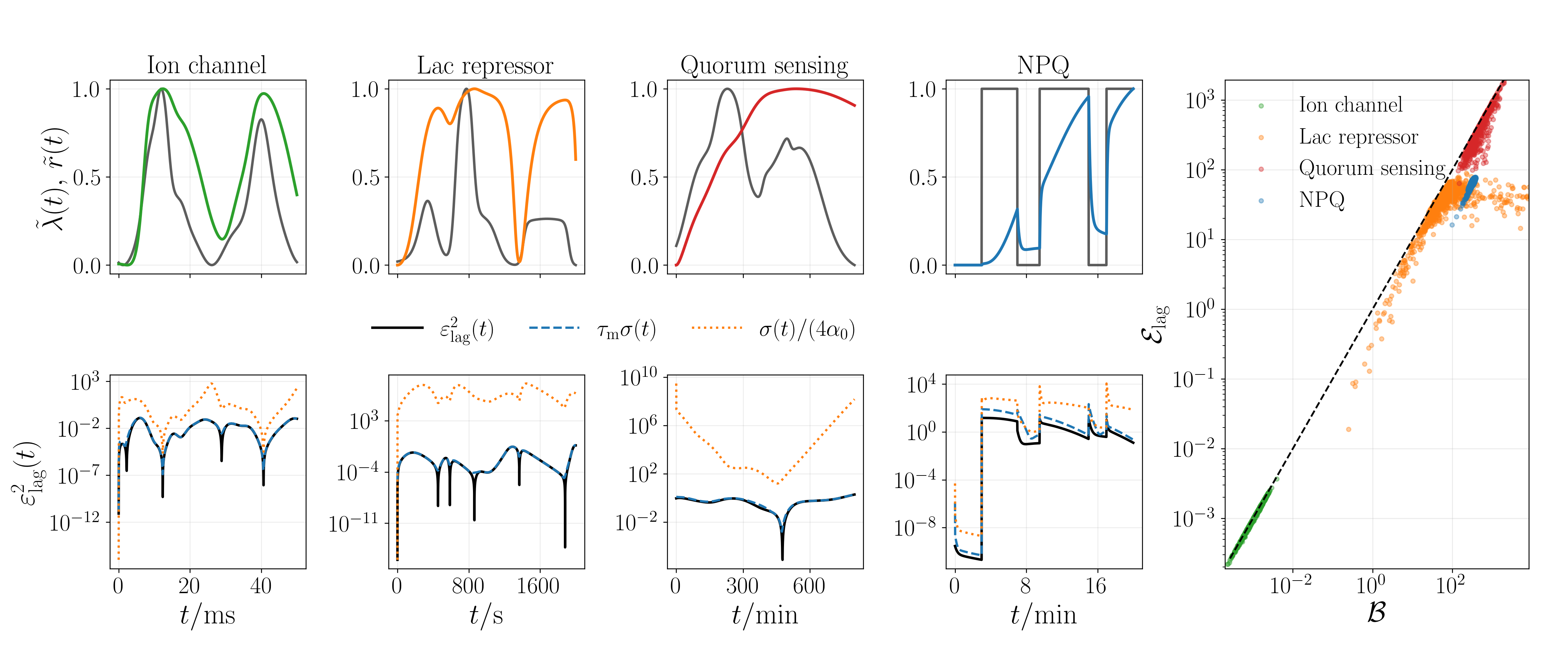}
\caption{
Thermodynamic lag-error bounds in four biochemical information-processing models. The first four columns show representative driven trajectories for the lac repressor, ligand-gated channel, quorum-sensing, and nonphotochemical quenching (NPQ) models. In the top row, black curves are normalized inputs $\tilde{\lambda}(t)\propto \lambda(t)$ and colored curves are normalized outputs $\tilde{r}(t)\propto \hat{r}(t)$. In the bottom row, black curves show the instantaneous squared lag error, blue dashed curves show the  bound
$\tau_\mathrm{m}\sigma$ with quadratic entropy production, and orange dotted curves show the Sobolev bound $\sigma/4\mlsc$ with the full entropy production. The
rightmost panel compares the integrated lag error with the integrated bound across 1000 random input protocols for each model. Across models and input ensembles, the lag error
remains below the quadratic thermodynamic bound, with the tightness depending on how strongly dissipation is projected onto output-relevant slow modes.
}
\label{fig:biochemical_bounds}
\end{figure*}

To test the generality of the thermodynamic bound, we analyzed four models of biological information processing spanning gene regulation, sensory transduction, collective signaling, and photoprotection: a Lac repressor induction model \cite{razo2018tuning}, a ligand-gated ion channel \cite{einav2017monod}, a receptor-level model of bacterial quorum sensing \cite{mehta2009information}, and a kinetic model of nonphotochemical quenching in photosynthetic organisms \cite{short2022xanthophyll}. For each system, an experimentally relevant input signal drives a reversible or locally reversible stochastic network, and the output corresponds to a biologically meaningful observable, such as promoter accessibility, channel opening probability, luminescence activation, or the concentration of active quenchers. In every case, the dynamics define an instantaneous frozen state that determines the equilibrium output and an associated lag error. The full and quadratic approximation to the entropy production and susceptibility can therefore be evaluated along the trajectory and compared with the lag error \cite{SM}.

Figure~\ref{fig:biochemical_bounds} shows representative input/output relationships, the lag errors, and their thermodynamic bounds for the four biochemical models. Details for the simulations are available in the SM. The inputs represent ligand or activator concentrations for the Lac repressor, ion channel, and quorum sensing models, and vary smoothly, while for the nonphotochemical quenching model the input represents the light intensity that varies discontinuously. In the ion channel and Lac repressor, the output is a relatively time local function of the input. As a consequence, the lag errors are small and tightly bounded by Eq. ~\ref{eq:main_bound}. For the quorum sensing and nonphotochemical quenching models, the input/output relationship manifests a larger memory effect with correspondingly larger lag error. Across the four models, the lag error is tightly constrained by the near equilibrium bound.  The  more general modified log-Sobolev inequality in Eq.~\ref{eq:sobolev_bound} is also shown, but very weakly bounds the lag error.  

The results of the biochemical models can be summarized by comparing the integrated lag error,
\begin{equation}
\mathcal E_{\mathrm{lag}}
=\int dt \,\epsilon_{\mathrm{lag}}^2(t),
\end{equation}
against the integrated thermodynamic bound,
\begin{equation}
\mathcal B =\int dt \,\tau_\mathrm{m}[\lambda(t)]\sigma(t)
\end{equation}
which is shown in Fig.~\ref{fig:biochemical_bounds} for 1000 random input time-series for each biochemical model. Comparisons to  Eq.~\ref{eq:sobolev_bound} are shown in the SM. 
Despite the substantial differences in network architecture, timescales, and biological function, all systems satisfy $\mathcal E_{\mathrm{lag}} \le \mathcal B$. The degree of saturation of the bound varies systematically among the models. Systems whose nonequilibrium response is dominated by a single slow mode exhibit nearly optimal use of dissipation and lie close to the theoretical limit. By contrast, systems with many competing relaxation pathways dissipate entropy in directions that do not efficiently contribute to the output, resulting in a looser bound. This behavior supports the interpretation of $\tau_\mathrm{m}$ as an equilibrium measure of computational memory.  Only dissipation projected onto output-relevant slow modes contributes effectively to reducing computational error. 

We have derived a thermodynamic bound on the accuracy of dynamical computation in reversible Markov networks \cite{wolpert2024stochastic}. The total computational error naturally separates into a representation error and a lag error, the latter arising from the inability of a stochastic system to instantaneously track a changing environment. We showed that the lag error is upper bounded by a product of the entropy production and an equilibrium measure of the memory stored in output-relevant modes. In this sense, computation requires not merely the expenditure of energy, but the selective conversion of dissipation into persistent memory of the quantities being computed. 

The bound establishes a general tradeoff between computational accuracy, memory, and dissipation. Accurate dynamical computation can be achieved either by dissipating substantial free energy or by storing information in slowly relaxing modes that retain memory of past inputs. These results place fundamental thermodynamic constraints on learning and adaptation in stochastic systems and provide a quantitative framework to understand the energetic costs of biological information processing \cite{hall2025entropy,tkavcik2025information} and the design of synthetic nonequilibrium computing architectures \cite{gao2021principles,chen2026optimal,freitas2026taming}.
The simulation code, and analysis scripts are available on Zenodo \cite{code}. 

\emph{Acknowledgments} This work was supported by NSF Grant No. CHE-2618387. GPT-5.6 Sol was used in manuscript preparation, the writing of simulation code, and the analysis of data, all with supervision and verification.

\bibliography{ref}

\end{document}


\title{Supporting Material: \\Entropy Production Bounds the Accuracy of Computation in Markov Networks}

\author{Songela W. Chen}
\affiliation{Department of Chemistry, University of California, Berkeley, CA, 94720, USA}
\affiliation{Kavli Energy NanoScience Institute, University of California, Berkeley, CA, 94720, USA}
\author{David T. Limmer}\email{dlimmer@berkeley.edu}
\affiliation{Department of Chemistry, University of California, Berkeley, CA, 94720, USA}
\affiliation{Kavli Energy NanoScience Institute, University of California, Berkeley, CA, 94720, USA}
\affiliation{Chemical Sciences Division, Lawrence Berkeley National Laboratory, Berkeley, CA 94720, USA}
\affiliation{Materials Sciences Division, Lawrence Berkeley National Laboratory, Berkeley, CA 94720, USA}

\maketitle 

\section{Relative Entropy and the Log-Sobolev Bound}

In this section we derive the upper bound on the lag error in terms of the entropy production and the modified log-Sobolev constant of the instantaneous generator.

\subsection{Bounding the lag by the relative entropy}

The lag error is defined as
\begin{equation}
\varepsilon_{\mathrm{lag}}= \mathbf{a}\cdot(\mathbf{p}-\boldsymbol{\pi})
\label{eq:S1}
\end{equation}
where the observable satisfies $a_i\in{0,1}$. Since
\begin{equation}
\left| \sum_i a_i(p_i-\pi_i) \right| \le \sum_i a_i |p_i-\pi_i| \le \sum_i |p_i-\pi_i|,
\end{equation}
we obtain
\begin{equation}
|\varepsilon_{\mathrm{lag}}| \le \frac12 \sum_i |p_i-\pi_i|, 
\label{eq:S2}
\end{equation}
where the factor of 1/2 follows because $a_i$ is a binary indicator function and therefore selects at most one side of the signed difference. Noting that the right hand side of Eq. \ref{eq:S2} is the total variation distance between instantaneous and frozen equilibrium distributions, Pinsker's inequality relates this distance to the relative entropy $D[\mathbf{p}||\boldsymbol{\pi}]$ between the distributions \cite{csiszar2011information,tsybakov_lower_2009}
\begin{equation}
\left ( \frac12 \sum_i |p_i-\pi_i| \right)^2 \le \frac12 D[p||\pi],
\label{eq:S3}
\end{equation}
where
\begin{equation}\label{eq:S4}
D[\mathbf{p}||\boldsymbol{\pi}]=\sum_i p_i \ln\frac{p_i}{\pi_i}\, .
\end{equation}
Combining Eqs.~\ref{eq:S2} and \ref{eq:S3} yields
\begin{equation}
\varepsilon_{\mathrm{lag}}^2 \le \frac12 D[p||\pi].
\label{eq:S5}
\end{equation}
Thus, controlling the lag error reduces to controlling the relative entropy from the instantaneous equilibrium distribution.

\subsection{Entropy production }
Since the frozen equilibrium distribution satisfies
\begin{equation}
\boldsymbol{\pi}\mathbf{W}=0, \qquad \pi_i k_{ij}=\pi_j k_{ji},
\end{equation}
the relative entropy is given by 
\begin{equation} 
D[\mathbf{p}||\boldsymbol{\pi}]=\sum_i p_i \ln \frac{p_i}{\pi_i}.
\label{eq:S6}
\end{equation}
For a fixed generator, differentiating with respect to time gives
\begin{align}
    \frac{dD}{dt} &= \sum_i \dot p_i \ln \frac{p_i}{\pi_i} + \dot p_i \\
    &= \sum_{i\neq j} \left( p_jk_{ji}- p_i k_{ij}\right) \ln \frac{p_i}{\pi_i}\\
    &=-\frac12 \sum_{i \neq j} (p_i k_{ij} - p_j k_{ji} ) \ln \frac{p_i k_{ij}}{p_j k_{ji}}
\end{align}
where the second equality invokes conservation of probability $\sum_i \dot p_i =0$ and substitutes the master equation, and the third equality invokes detailed balance and symmetrizes over $i$ and $j$~\cite{esposito_three_2010}.
We therefore identify the total entropy production rate
\begin{equation}
\sigma = \frac12 \sum_{i \neq j} (p_i k_{ij} - p_j k_{ji} ) \ln \frac{p_i k_{ij}}{p_j k_{ji}}
\label{eq:full-entropy}
\end{equation}
such that
\begin{equation}
\frac{d}{dt}D[\mathbf{p}||\boldsymbol{\pi}]=-\sigma.
\label{eq:S8}
\end{equation}
This matches the Schnakenberg formula typically seen in stochastic thermodynamics \cite[Eq. 3.39]{peliti2021stochastic}, \cite[Section 9.2]{limmer2024statistical}.
Since
\begin{equation}
(x-y)\ln(x/y)\ge 0,
\end{equation}
the entropy production is nonnegative and the relative entropy decreases monotonically toward equilibrium for the frozen generator. The entropy production can be rewritten as
\begin{equation}
    \sigma = \frac12 \sum_{ij}\pi_i k_{ij} (f_i-f_j) \ln \frac{f_i}{f_j}\, ,
\end{equation}
introducing $f_i=p_i/\pi_i$, which will be useful for subsequent relationships. 

\subsection{Modified log-Sobolev inequality}

The family of logarithmic Sobolev (``log-Sobolev") inequalities measure relaxation to equilibrium for Markov chain dynamics \cite{gross_logarithmic_1975,diaconis1996logarithmic,bobkov_modified_2006}. Ref. \cite{bobkov_modified_2006} explores a modified log-Sobolev inequality
\begin{equation}
    \mlsc \mathrm{Ent}_{\pi}(f)\leq \frac{1}{2}\mathcal{E}(f, \ln f)
    \label{eq:modified-log-sobolev-bobkov}
\end{equation}
where
\begin{equation}
\mathcal E(f,g) =\frac12 \sum_{ij} \pi_i k_{ij} (f_i-f_j)(g_i-g_j)
\label{eq:dirichlet}
\end{equation}
is the Dirichlet form for the reversible Markov kernel 
\cite[Example 3.4]{bobkov_modified_2006}
and
\begin{equation}
\mathrm{Ent}_{\pi}(f)= \sum_i \pi_i f_i\ln f_i-\left( \sum_i \pi_i f_i\right) \ln \left(\sum_i \pi_i f_i\right)
\end{equation}
is the entropy functional on $f$. Given our choice of $f=\bf{p}/\bm{\pi}$ and conservation of probability,
\begin{equation}
\sum_i \pi_i f_i= \sum_i p_i= 1,
\end{equation}
we have
\begin{equation}
\mathrm{Ent}_{\pi}(f) =D[p||\pi].
\end{equation}
Furthermore, the Dirichlet form here corresponds to the rate of entropy production
\begin{equation}
\mathcal E(f,\ln f) =\sigma\, .
\end{equation}
The modified log-Sobolev inequality (Eq. \ref{eq:modified-log-sobolev-bobkov}) therefore implies
\begin{equation}
D[\mathbf{p}||\boldsymbol{\pi}] \le \frac{\sigma}{2\mlsc} \, ,
\label{eq:S10}
\end{equation}
providing a bound on the rate of decay of the informational entropy through the modified log-Sobolev constant $\mlsc$.
\footnote{Since $\sigma = -\frac{d}{dt}D[\mathbf{p}||\boldsymbol{\pi}]$, then $\frac{d}{dt}D[\mathbf{p}||\boldsymbol{\pi}] \leq -2 \mlsc D[\mathbf{p}||\boldsymbol{\pi}] $. An associated relation \cite[Theorem 2.4]{bobkov_modified_2006} gives the decay explicitly as a function of time, $D[p(t)||\pi] \leq  D[p(0)||\pi]e^{-2\mlsc t}$.}
Combining Eqs.~\ref{eq:S5} and \ref{eq:S10} gives the thermodynamic estimate
\begin{equation}
\varepsilon_{\mathrm{lag}}^2\le \frac{\sigma}{4\mlsc}.
\label{eq:S11}
\end{equation}
Equation \ref{eq:S11} is exact and holds for arbitrary deviations from equilibrium. 

\subsection{Estimates of the modified log-Sobolev constant}
The bound
\begin{equation}
\varepsilon_{\mathrm{lag}}^2 \le \frac{\sigma}{4\mlsc}
\end{equation}
depends on the modified log-Sobolev constant $\mlsc$, which is generally difficult to compute exactly. In practice, several estimates are available. For a reversible continuous-time Markov chain, $\mlsc$ may be solved numerically by constrained optimization over positive functions $f$, since $\mlsc$ is defined variationally by Eq. \ref{eq:modified-log-sobolev-bobkov}
\begin{equation}
    \mlsc \leq \frac{\mathcal E(f,\ln f)}{ 2\mathrm{Ent}_{\pi}(f) }\, .
    \label{eq:mlsc-variational}
\end{equation} For the small biochemical models considered in the main text, this procedure provides an essentially exact value of $\mlsc$.

A useful lower bound can be obtained from the spectral gap. Let
\begin{equation}
0=\omega_0<\omega_1\le\omega_2\le\cdots
\label{eq:spectrum}
\end{equation}
be the eigenvalues of -$\mathbf{W}$, and denote the spectral gap by $\omega=\omega_1$. For finite reversible Markov chains, the classical log-Sobolev constant $\alpha$ satisfies \cite[Thm. A.1 and Corr. A.4]{diaconis1996logarithmic}
\begin{equation}
\alpha \geq \frac{(1 - 2 \pi_{\min}) \omega }{\ln[1/\pi_{\min}-1]} \ge \frac{\pi_{\min}}{2} \omega ,
\label{eq:S13}
\end{equation}
where $\pi_{\min}=\min_i \pi_i$.
This estimate is useful because both $\omega$ and $\pi_{\min}$ are easily computed from the instantaneous generator. The modified log-Sobolev constant further satisfies \cite[Prop. 3.6]{bobkov_modified_2006}
\begin{equation}
    \mlsc \geq \alpha \ge \frac{\pi_{\min}\omega}{2} \, .
    \label{eq:mlsc-estimate}
\end{equation}
Combining Eqs.~\ref{eq:S11}, \ref{eq:S13}, and \ref{eq:mlsc-estimate} gives
\begin{equation}
\varepsilon_{\mathrm{lag}}^2 \le \frac{\sigma}{ 2\pi_{\min}\omega},
\label{eq:S15}
\end{equation}
which provides a completely explicit, though generally conservative, estimate of the lag error.

For systems whose dynamics are dominated by a single slow relaxation mode, one expects $\mlsc \sim \omega$. This observation explains why the spectral-gap estimate often performs substantially better in practice than the bound Eq. \ref{eq:S13} would suggest.
Further, the appearance of the spectral gap indicates that slowly relaxing modes simultaneously increase the memory time of the network and decrease the modified log-Sobolev constant. Consequently, systems with long-lived memory can exhibit large lag errors while also possessing the smallest entropy-production rates required to sustain them.

\section{Derivation of the Quadratic Thermodynamic Bound}
In this section we derive the quadratic bound on the lag error that forms the central result of the main text. Unlike the modified log-Sobolev bound, the present result is asymptotically exact only in the regime where the probability distribution remains close to the instantaneous equilibrium state.

We parameterize small deviations from the frozen equilibrium distribution by
\begin{equation}
p_i=\pi_i(1+h_i),
\label{eq:S16}
\end{equation}
where
\begin{equation}
h_i=\frac{p_i-\pi_i}{\pi_i}.
\end{equation}
Normalization of the probability distribution implies
\begin{equation}
\sum_i \pi_i h_i=0.
\label{eq:S17}
\end{equation}
Expanding the relative entropy,
\begin{equation}
D[\mathbf{p}||\boldsymbol{\pi}] = \sum_i \pi_i(1+h_i) \ln(1+h_i),
\end{equation}
gives
\begin{equation}
D[\mathbf{p}||\boldsymbol{\pi}] =\frac12 \sum_i \pi_i h_i^2 + O(h^3).
\label{eq:S18}
\end{equation}
Similarly,
\begin{equation}
f_i =1+h_i, \qquad \ln f_i=h_i +O(h^2),
\end{equation}
so that the entropy production becomes
\begin{equation}
\sigma = \frac12 \sum_{ij} \pi_i k_{ij} (h_i-h_j)^2 +O(h^3)
\label{eq:quad-entropy}
\end{equation}
for small deviations from equilibrium.

The lag error can be written as
\begin{equation}
\varepsilon_{\mathrm{lag}}= \sum_i a_i(p_i-\pi_i) =\sum_i \pi_i \bar a_i h_i,
\label{eq:S23}
\end{equation}
where
\begin{equation}
\bar a_i = a_i-\langle a\rangle_\pi
\end{equation}
is the centered observable. We now introduce a function $\mathbf{q}$ satisfying the discrete Poisson equation
\begin{equation}
\mathbf{W} \mathbf{q}=-\bar{\mathbf{a}} , \qquad  \mathbf{q} \cdot \boldsymbol{{\pi}}=0.
\label{eq:poisson}
\end{equation}
Since $\langle \bf{\bar a}\rangle_\pi=0$, Eq.~\ref{eq:poisson} possesses a unique solution on the subspace orthogonal to the stationary distribution. Component-wise
\begin{equation}
-\sum_j k_{ij}(q_j-q_i)=\bar a_i
\label{eq:S25}
\end{equation}
which follows from conservation of probability on $\bf{W}$. Substituting Eq.~\ref{eq:S25} into Eq.~\ref{eq:S23} yields
\begin{equation}
\varepsilon_{\mathrm{lag}}=\frac12 \sum_{ij} \pi_i k_{ij} (q_i-q_j)(h_i-h_j).
\label{eq:S26}
\end{equation}
Applying the Cauchy-Schwarz inequality to Eq.~\ref{eq:S26} gives
\begin{align}
\varepsilon_{\mathrm{lag}}^2 \le& \left[\frac12  \sum_{ij} \pi_i k_{ij} (q_i-q_j)^2 \right] \times \left[\frac12  \sum_{ij} \pi_i k_{ij} \left (h_i-h_j\right)^2 \right].
\label{eq:S27}
\end{align}
Using the quadratic form for the entropy production we obtain
\begin{equation}
\varepsilon_{\mathrm{lag}}^2 \le \tau_\mathrm{m} \sigma 
\label{eq:S28}
\end{equation}
where
\begin{equation}
\tau_\mathrm{m}=\frac12  \sum_{ij} \pi_i k_{ij} (q_i-q_j)^2
\label{eq:S29}
\end{equation}
represents an observable susceptibility. Equation \ref{eq:S28} is asymptotically exact for small deviations from the frozen equilibrium state and constitutes the principal result of this work.

Equality in Eq.~\ref{eq:S28} occurs when the two vectors appearing in the Cauchy-Schwarz inequality are proportional 
\begin{equation}
h_i - h_j = \Lambda (q_i-q_j) \qquad \forall (i,j)
\ \text{with}\ k_{ij}>0 .
\end{equation}
Equivalently,
\begin{equation}
h_i= \Lambda q_i + {\rm const}.
\end{equation}
Because Eq.~\ref{eq:S17} and $\mathbf{q} \cdot \boldsymbol{\pi}=0$ remove the additive constant, the equality condition becomes
\begin{equation}
p_i-\pi_i= \Lambda  \pi_i q_i.
\label{eq:S30}
\end{equation}
Thus the bound becomes tight precisely when the nonequilibrium displacement of the probability distribution is aligned with the output committor $\mathbf{q}$. In this case, all of the dissipation is stored in the output-relevant relaxation mode and is therefore used optimally to reduce computational error.

\section{Properties of the integrated susceptibility  $\tau_\mathrm{m}$}
The integrated susceptibility
\begin{equation}
\tau_\mathrm{m}=\frac12 \sum_{ij} \pi_i k_{ij} (q_i-q_j)^2
\label{eq:S31}
\end{equation}
plays a central role in the quadratic thermodynamic bound. In this section we derive several equivalent representations of $\tau_\mathrm{m}$ and discuss its interpretation. 

First, $\tau_\mathrm{m}$ can be expressed as the Dirichlet form of the reversible Markov kernel (Eq. \ref{eq:dirichlet})
\begin{equation}
\tau_\mathrm{m}=\mathcal E(\mathbf{q},\mathbf{q}).
\label{eq:S32}
\end{equation}
Define the inner product over the stationary distribution as 
\begin{equation}
\langle f,g\rangle_\pi=\sum_i \pi_i f_i g_i.
\end{equation}
Using the discrete Poisson equation \ref{eq:poisson}
together with the identity \cite[Eq. 1.1]{bobkov_modified_2006}
\begin{equation}
\mathcal E(f,g)=-\langle f,Wg\rangle_\pi ,
\end{equation}
we obtain
\begin{equation}
\tau_\mathrm{m}= -\langle \mathbf{q},\mathbf{W} \mathbf{q}\rangle_\pi = \langle \mathbf{q},\bar {\mathbf{a}}\rangle_\pi 
\label{eq:Ca_inner}
\end{equation}
hence,
\begin{equation}
\tau_\mathrm{m}=  \sum_i \pi_i q_i\bar a_i .
\label{eq:S33}
\end{equation}
Because
$\langle \bar{\mathbf{a}}\rangle_\pi=0$,
the Poisson equation may be inverted on the subspace orthogonal to the stationary distribution,
\begin{equation}
\mathbf{q}=-\mathbf{W}^{-1}\bar{\mathbf{a}}.
\label{eq:poisson_invert}
\end{equation}
Equivalently,
\begin{equation}
\mathbf{q}=\int_0^\infty dt e^{\mathbf{W}t}\bar{\mathbf{a}} ,
\label{eq:S34}
\end{equation}
where the integral converges because all nonzero eigenvalues of $\mathbf{W}$ are negative.
Substituting Eq.~\ref{eq:S34} into Eq.~\ref{eq:Ca_inner} yields
\begin{align}
\tau_\mathrm{m} &=  \int_0^\infty dt \, \langle \bar {\mathbf{a}}, e^{\mathbf{W}t}\bar {\mathbf{a}} \rangle_\pi
\nonumber\\
&= \int_0^\infty dt \, \langle \bar {\mathbf{a}}(0)\bar {\mathbf{a}}(t) \rangle .
\label{eq:Ca_autocorrelation}
\end{align}
Equation \ref{eq:Ca_autocorrelation} shows that $\tau_\mathrm{m}$ is the integrated equilibrium autocorrelation time of the output observable.

Next, define the orthonormal basis expansion of the Markov kernel as 
\begin{equation}
\mathbf{W}\boldsymbol{\phi}_n=-\omega_n \boldsymbol{\phi}_n.
\end{equation}
Expanding the centered observable in this basis,
\begin{equation}
\bar {\mathbf{a}} = \sum_{n>0} c_n \boldsymbol{\phi}_n, \qquad c_n=\langle \boldsymbol{\phi}_n,\bar {\mathbf{a}}\rangle_\pi,
\end{equation}
gives via the inverted Poisson equation \ref{eq:poisson_invert},
\begin{equation}
\mathbf{q} = \sum_{n>0} \frac{c_n}{\omega_n} \boldsymbol{\phi}_n .
\end{equation}
Substituting into Eq.~\ref{eq:S33} yields
\begin{equation}
\tau_\mathrm{m}=  \sum_{n>0} \frac{c_n^2}{\omega_n}.
\label{eq:susceptibility_eigenbasis}
\end{equation}
The susceptibility is therefore dominated by slowly relaxing modes that overlap strongly with the observable.

Since the spectral gap $\omega$ is defined by the dominant nonzero eigenvalue (Eq. \ref{eq:spectrum}),
Eq.~\ref{eq:susceptibility_eigenbasis} immediately implies
\begin{equation}
\tau_\mathrm{m}
\le
\frac{1}{\omega}
\sum_{n>0}
c_n^2 .
\end{equation}
Using Parseval's identity~\cite[Section 6B]{axler_linear_2024},
\begin{equation}
\sum_{n>0}
c_n^2= \langle \bar {\mathbf{a}}^2 \rangle_\pi
\end{equation}
we obtain
\begin{equation}
\tau_\mathrm{m}\le \frac{\langle \bar {\mathbf{a}}^2 \rangle_\pi}{\omega}.
\label{eq:S37}
\end{equation}
For binary observables, if the occupancies of the $n$ output sites are uncorrelated, then
\begin{equation}
\langle \bar {\mathbf{a}}^2 \rangle_\pi=\left [\langle a\rangle_\pi \left(1-\langle a\rangle_\pi \right)\right ]^n
 \le \left (\frac14  \right )^n,
\end{equation}
and therefore
\begin{equation}
\tau_\mathrm{m} \le \frac{1}{\omega} \left (\frac14  \right )^n 
\label{eq:S38}
\end{equation}
which shrinks with increasing $n$.

Equation \ref{eq:Ca_autocorrelation} shows that $\tau_\mathrm{m}$ is an equilibrium memory time associated with the output observable. Equation \ref{eq:susceptibility_eigenbasis} further demonstrates that only relaxation modes with both (i) long relaxation times and (ii) significant overlap with the observable, contribute appreciably to the susceptibility. Consequently, the quadratic thermodynamic bound may be interpreted as stating that entropy production can reduce computational error only insofar as it is stored in slow, output-relevant modes of the dynamics. Dissipation occurring in modes orthogonal to the observable contributes to entropy production but does not improve computational performance.

\section{Artificial Learning Models}
This section describes the synthetic reversible Markov networks used in the numerical experiments of the main text. The network consists of $N$ states arranged on a ring with nearest- and third-nearest-neighbor connections,
\begin{equation}
i  \leftrightarrow i+1 \pmod N, \qquad i \leftrightarrow i+3 \pmod N.
\label{eq:S39}
\end{equation}
Unless otherwise stated, we use $N=10$. The output observable is defined on two designated states (rounding down when necessary),
\begin{equation}
a_i= \begin{cases} 1, & i=N/3, 2N/3 \\
0, & \text{otherwise}.
\end{cases}
\label{eq:S40}
\end{equation}
The network therefore computes a scalar output, $\hat r(t) = \mathbf{a} \cdot  \mathbf{p}(t)$.

The training and testing input trajectories are generated from
\begin{align}
\lambda(t)= 0.50 + 0.30\sin(st+\phi_1)+0.16\sin(2.3st+\phi_2)+0.08\sin(4.1st+\phi_3),
\label{eq:S46}
\end{align}
where the phases
\begin{equation}
\phi_1,\phi_2,\phi_3 \sim{\rm Uniform}(0,2\pi)
\end{equation}
are drawn independently. The resulting signal is clipped to
\begin{equation}
0.02 \le \lambda(t) \le 0.98.
\end{equation}
The parameter $s$ controls the driving speed and therefore the magnitude of nonequilibrium lag.

Network site energies are taken to depend affinely on the input,
\begin{equation}
E_i(\lambda)=E_{0,i}+E_{1,i}\lambda ,
\label{eq:S41}
\end{equation}
while the edge barriers satisfy
\begin{equation}
g_e(\lambda) =g_{0,e} + g_{1,e}\lambda .
\label{eq:S42}
\end{equation}
For an edge $e=(ij)$, the transition rates are
\begin{equation}
k_{ij}(\lambda)= \exp \left[ g_{ij}(\lambda)+\frac\beta2 \left( E_i(\lambda)-E_j(\lambda)\right)\right],
\label{eq:S43}
\end{equation}
and
\begin{equation}
k_{ji}(\lambda)=\exp\left[g_{ij}(\lambda)+\frac\beta2 \left( E_j(\lambda)-E_i(\lambda)\right)\right].
\label{eq:S44}
\end{equation}
which satisfy detailed balance with respect to
\begin{equation}
\pi_i(\lambda)=\frac{e^{-\beta E_i(\lambda)}}{\sum_j e^{-\beta E_j(\lambda)}}.
\label{eq:S45}
\end{equation}
Because the transition rates depend exponentially on both the node energies $E_i$ and edge barriers $g_{ij}$, we may train the network to express arbitrary nonlinear behaviors.

Two classes of target computations were considered, a memoryless target and a history-dependent one. The target output for the memoryless target is a single-valued function of the instantaneous input,
\begin{equation}
r(t)=f[\lambda(t)]\, , \qquad f[\lambda]=0.04+\frac{0.88}{1+\exp[-8(\lambda-0.52)]}
\label{eq:S47}
\end{equation}
while for the history-dependent target, the output obeys
\begin{equation}
\tau(t)\dot r=f[\lambda(t)]- r(t),
\label{eq:S49}
\end{equation}
with asymmetric relaxation time
\begin{equation}
\tau(t)=\begin{cases} 0.55, & f[\lambda(t)]>r(t), \\
1.45, & f[\lambda(t)]\le r(t).
\end{cases}
\label{eq:S50}
\end{equation}
The latter target depends explicitly on the past trajectory of the input and therefore requires memory. In both cases, we train the network to express these input-output mappings and measure errors relative to the analytical output $r(t)$ prescribed by the given input $\lambda(t)$.

Training was performed in two stages. In the first stage, the energy parameters $\{E_{0,i},E_{1,i}\}$
were optimized to minimize the representation loss,
\begin{equation}
\mathcal L_{\rm rep}= \frac{1}{T} \int_0^T dt, \left[ r(t)-r_{\rm eq}(t) \right]^2,
\label{eq:S51}
\end{equation}
where $r_{\rm eq}(t) = \mathbf{a} \cdot  \bm{\pi}[\lambda(t)]$ is associated with the frozen equilibrium distribution for $\lambda(t)$ at a given time.
In the second stage,  all parameters, $\{E_{0,i},E_{1,i},g_{0,e},g_{1,e}\},$
were optimized to minimize the dynamical loss,
\begin{equation}
\mathcal L_{\rm dyn}=\frac{1}{T}\int_0^T dt, \left[ r(t)-\hat r(t)\right]^2 
\label{eq:S52}
\end{equation}
where $\hat r(t) = \mathbf{a} \cdot  \mathbf{p}(t)$ is computed from the instantaneous probability density.
To avoid pathologically large or small transition rates, the barriers were regularized according to
\begin{equation}
\mathcal L =\mathcal L_{\rm dyn} +\lambda_g \sum_e (g_e-g^\star)^2,
\label{eq:S53}
\end{equation}
where $g^\star$ is a prescribed reference barrier. Optimization was performed using the Adam optimizer in PyTorch with gradients computed by automatic differentiation through the master equation integrator. For each target class, a batch of 24 input-output mappings with varying speeds and phases were used.

For each trained network we computed
\begin{align}
\mathcal E_{\rm lag}&= \int_0^T dt \, \varepsilon_{\rm lag}^2(t),\\
\mathcal E_{\rm rep}&=\int_0^T dt \, \varepsilon_{\rm rep}^2(t),
\\
\mathcal B &= \int_0^T dt \, \tau_\mathrm{m}[\lambda(t)] \sigma(t).
\end{align}
The susceptibility $\tau_\mathrm{m}$ was obtained by solving the Poisson equation
\begin{equation}
\mathbf{W}[\lambda(t)] \mathbf{q}(t)=-\bar{\mathbf{a}}(t)
\end{equation}
at each time point and evaluating Eq.~\ref{eq:S31}. 

\begin{figure}
    \centering
    \includegraphics[width=.5\linewidth]{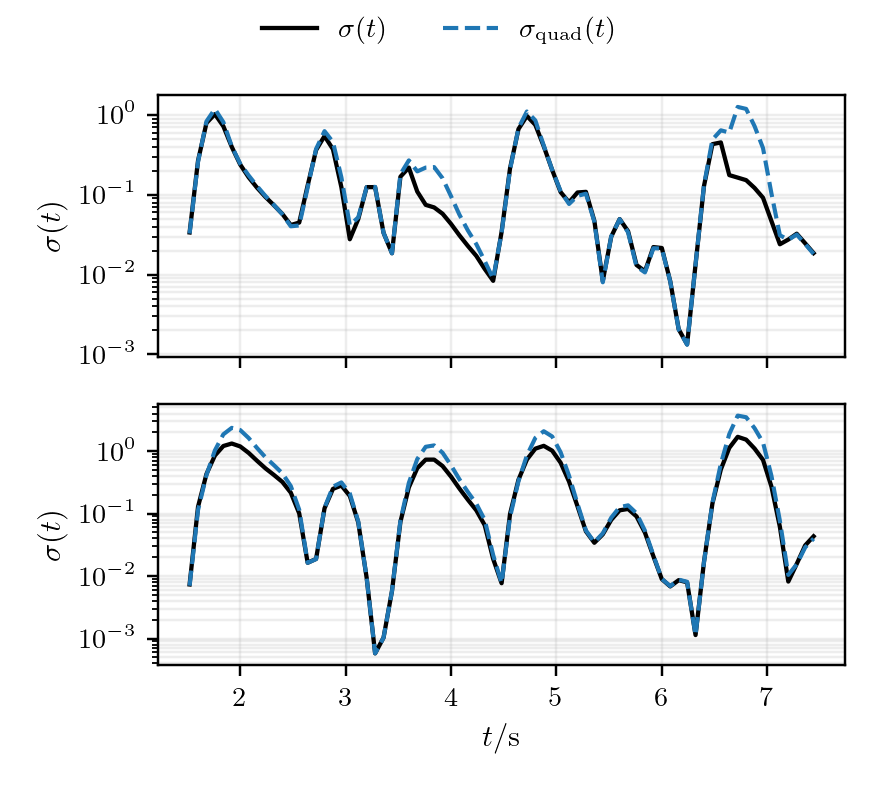}
\caption{
Comparison between the near equilibrium and full calculation of the entropy production  for trained reversible Markov networks with $N=10$ states. The upper panel shows a memoryless target. The lower panel shows a history-dependent target. 
}
\label{fig:entropy-trained}
\end{figure}

The main text demonstrates that the bound derived from the Cauchy-Schwarz inequality tightly constrains the lag error in both target models. In Fig.~\ref{fig:entropy-trained} we confirm that over both time-series, the quadratic approximation to the entropy production is a reasonable approximation to the full entropy production. There are some deviations at local maxima of the entropy production in both target models, where nodes of the Markov model pass through configurations with small probability, leading to over-estimates of $\sigma$.

\section{Biological Information Processing Models}
This section describes the biological information-processing models used to test the generality of the thermodynamic bounds derived in the main text. We use two models involving simple ligand concentration inputs, the nicotinic acetylcholine receptor and the Lac repressor. As an example of multiple inputs, we model quorum sensing in bacteria. Finally, a more complex non-linear time-dependent input is exhibited through non-photochemical quenching in photosystem II.
While these models are not intended to be quantitatively faithful to their experimental biochemical sources, they represent a range of different behaviors that demonstrate the generality of the thermodynamic bounds we derive.

\subsection{Ligand-Gated Ion Channel}

The nicotinic acetylcholine receptor (nAChR) is a ligand-gated ion channel, where acetylcholine binding at two sites induces ion channel opening. 
As a starting point we use the 
Monod–Wyman–Changeux (MWC) model analyzed in Ref.~\cite{einav2017monod}, but we ignore the states where the channel is open with fewer than two ligands bound since they have low probability. Our reduced model obeys detailed balance for the remaining four states:
\begin{equation}
C_0
\underset{k_{\rm off}}
{\overset{2k_{\rm on}\lambda(t)}{\rightleftharpoons}}
C_1
\underset{2k_{\rm off}}
{\overset{k_{\rm on}\lambda(t)}{\rightleftharpoons}}
C_2
\underset{\nu^{-1}}
{\overset{\nu}{\rightleftharpoons}}
O .
\end{equation}
Here \(C_0\), \(C_1\), and \(C_2\) denote closed states with zero, one, and
two bound ligands, while \(O\) is the open state. The factors of two in the
first binding and second unbinding reactions account for the degeneracy of two
equivalent ligand-binding sites. The input is the acetylcholine (ACh) concentration,
\begin{equation}
\lambda(t)=[{\rm ACh}](t),
\end{equation}
and the output is the open probability,
\begin{equation}
r(t)=p_O(t).
\end{equation}
Because the state graph is a line, detailed balance is automatically satisfied
for each fixed value of \(\lambda\). The associated frozen equilibrium is defined by the MWC state weights~\cite[Fig. 2A and Eq. 1]{einav2017monod}, which also define the detailed-balance rates. 
Defining
\begin{equation}
x(\lambda)=\frac{k_{\rm on}\lambda}{k_{\rm off}},
\end{equation}
the unnormalized stationary weights are
\begin{equation}
w_{C_0}=1,\qquad
w_{C_1}=2x,\qquad
w_{C_2}=x^2,\qquad
w_O=x^2\frac{\nu}{\nu^{-1}}.
\end{equation}
Thus
\begin{equation}
\pi_O(\lambda)=
\frac{w_O}{\sum_i w_i} = 
\frac{x^2\nu/\nu^{-1}}
{1+2x+x^2(1+\nu/\nu^{-1})}.
\end{equation}
This is equivalent to Eq. 1 of Ref.~\cite{einav2017monod} ignoring the 0- and 1-ligand-bound open states.

The numerical parameters are derived from the
wild-type nAChR MWC parameterization of Ref.~\cite{einav2017monod}. First, we use the MWC
closed-state ligand dissociation constant,
\begin{equation}
K_C=\frac{k_{\rm off}}{k_{\rm on}}=60\,\mu{\rm M}.
\end{equation}
and use a textbook ligand binding rate $k_\mathrm{on} = 5 \times 10^7 \, \mathrm{M}^{-1}\mathrm{s}^{-1}$ \cite{changeux2005nicotinic}.
Second, the opening-to-closing ratio $\nu/\nu^{-1}$ is derived from the MWC state weights
\begin{equation}
\frac{\nu}{\nu^{-1}}
=
e^{\beta\epsilon}
\left(\frac{K_C}{K_O}\right)^2,
\end{equation}
where \(K_C=60\,\mu{\rm M}\), \(K_O=0.1\,{\rm nM}\) is the open-state ligand dissociation constant, and
\(\beta\epsilon=-23.7\) denotes $\epsilon$ as the energy to go from open to closed conformations without ligand bound. While the open-state ligand dissociation constant $K_O$ does not define direct transitions in our model, it factors into the opening-to-closing equilibrium phenomenologically. This gives
\begin{equation}
\frac{\nu}{\nu^{-1}}=18.345.
\end{equation}
We set the closing rate to \(\nu^{-1}=10^3\,\mathrm{s}^{-1}\) based on a millisecond timescale for a channel to stay open before closing again~\cite[pg. S4]{einav2017monod}.
The parameters are summarized in Table \ref{tab:ligand_channel_parameters}.

\begin{table}[h]
\centering
\begin{tabular}{lll}
\toprule
Parameter & Value & Units \\
\midrule
\(K_C=k_{\rm off}/k_{\rm on}\) & \(60\times10^{-6}\) & \(\mathrm{M}\) \\
\(k_{\rm on}\) & {\(5.0\times10^7\)} & \(\mathrm{M}^{-1}\mathrm{s}^{-1}\) \\
\(k_{\rm off}\) & \(3.0\times10^3\) & \(\mathrm{s}^{-1}\) \\
\(\nu/\nu^{-1} = e^{\beta\epsilon}
\left(\frac{K_C}{K_O}\right)^2\) & \(18.345\) & dimensionless \\
\(\nu^{-1}\) & \(1.0\times10^3\) & \(\mathrm{s}^{-1}\) \\
\(\nu\) & \(1.8345\times10^4\) & \(\mathrm{s}^{-1}\) \\
\bottomrule
\end{tabular}
\caption{Rate parameters for the reduced ligand-gated ion channel model. The
ligand equilibrium constants and gate open-closed energy change are taken from
wild-type nAChR MWC parameters of Ref.~\cite{einav2017monod}.}
\label{tab:ligand_channel_parameters}
\end{table}

\subsection{Lac Repressor Induction}
We use a minimal model of allosteric induction in the Lac repressor \cite{razo2018tuning}. When active and bound to the promoter, the Lac repressor prevents DNA transcription. However, the Lac repressor can be inactivated allosterically by its inducer allolactose. Like for the acetylcholine receptor, our model is adapted from the Monod-Wyman-Changeux (MWC) model in Ref. \cite{razo2018tuning}, where probabilities of being in the various states are converted into detailed-balanced transition rates. 
The state of the model is
\begin{equation}
X=(\chi,s,b),
\end{equation}
where \(\chi\in\{A,I\}\) denotes the active or inactive Lac repressor
conformation, \(s\in\{0,1,2\}\) is the number of bound IPTG molecules (allolactose mimic used in experiment), and
\(b\in\{F,B\}\) denotes whether the DNA promoter is free or bound by repressor.
The full state space therefore contains \(2\times 3\times 2=12\) states,
\begin{equation}
\{A_0F,A_1F,A_2F,I_0F,I_1F,I_2F,
  A_0B,A_1B,A_2B,I_0B,I_1B,I_2B\}.
\end{equation}
The input signal is the IPTG concentration,
\begin{equation}
\lambda(t)=[\mathrm{IPTG}](t),
\end{equation}
and the output is the transcriptionally available (free) promoter fraction,
\begin{equation}
r(t)=\sum_{\chi\in\{A,I\}}\sum_{s=0}^{2}p_{\chi s F}(t).
\end{equation}

For a fixed value of \(\lambda\), the target equilibrium distribution is
defined by the MWC weights
\begin{align}
w_{A s F}(\lambda)
&=\binom{2}{s}\left(\frac{\lambda}{K_A}\right)^s,\\
w_{I s F}(\lambda)
&=e^{-\Delta\epsilon_{AI}}
  \binom{2}{s}\left(\frac{\lambda}{K_I}\right)^s,\\
w_{A s B}(\lambda)
&=B_A\binom{2}{s}\left(\frac{\lambda}{K_A}\right)^s,\\
w_{I s B}(\lambda)
&=B_I e^{-\Delta\epsilon_{AI}}
  \binom{2}{s}\left(\frac{\lambda}{K_I}\right)^s, 
\end{align}
whose associated stationary distribution is
\begin{equation}
\pi_{\chi s b}(\lambda)=
\frac{w_{\chi s b}(\lambda)}
{\sum_{\chi',s',b'}w_{\chi' s' b'}(\lambda)}.
\end{equation}
Here \(K_A\) and \(K_I\) are the IPTG dissociation constants for the active
and inactive conformations, respectively, and \(\Delta\epsilon_{AI}\) is the
inactive-active conformational free-energy difference in units of \(k_\mathrm{B} T\).
The dimensionless active-state DNA-binding factor is
\begin{equation}
B_A=\frac{R}{N_{NS}}e^{-\Delta\epsilon_{RA}},
\end{equation}
where \(R\) is the LacI dimer copy number, \(N_{NS}\) is the number of
nonspecific genomic binding sites, and \(\Delta\epsilon_{RA}\) is the
active-repressor operator binding energy. The model in Ref. \cite{razo2018tuning} neglects binding of
the inactive repressor to the operator. In our model
we retain a small finite inactive binding factor \(B_I\) so that the inactive
promoter-bound states remain present for model stability but weakly populated.

With the definitions
\begin{equation}
W_A(\lambda)=\left(1+\frac{\lambda}{K_A}\right)^2,
\qquad
W_I(\lambda)=e^{-\Delta\epsilon_{AI}}
\left(1+\frac{\lambda}{K_I}\right)^2,
\end{equation}
the frozen equilibrium output can be written as
\begin{equation}
r_{\rm eq}(\lambda)=
\frac{W_A(\lambda)+W_I(\lambda)}
{[1+B_A]W_A(\lambda)+[1+B_I]W_I(\lambda)}.
\end{equation}
In the limit \(B_I\to0\), this reduces to the simple-repression MWC
fold-change expression used in Ref.~\cite[Eq. 4-5]{razo2018tuning},
\begin{equation}
r_{\rm eq}(\lambda)=
\left[1+B_A p_A(\lambda)\right]^{-1},
\qquad
p_A(\lambda)=
\frac{W_A(\lambda)}{W_A(\lambda)+W_I(\lambda)}.
\end{equation}
We reproduce the fold-change characteristics from Ref. \cite{razo2018tuning} with the inactive binding population present.

The model has three classes of reversible transitions.
First, ligand binding and unbinding occur independently on the two ligand-binding
sites. For $\chi\in \{A,I\}$,
\begin{equation}
\chi_0
\underset{k_\mathrm{off}^\chi}
{\overset{2k_\mathrm{on}^\chi\lambda(t)}{\rightleftharpoons}}
\chi_1
\underset{2k_\mathrm{off}^\chi}
{\overset{k_\mathrm{on}^\chi\lambda(t)}{\rightleftharpoons}}
\chi_2
\end{equation}
where like before the factor of 2 accounts for degeneracy in the two ligand binding sites, with equilibrium constants
\begin{equation}
K_A=\frac{k_\mathrm{off}^A}{k_\mathrm{on}^A},
\qquad
K_I=\frac{k_\mathrm{off}^I}{k_\mathrm{on}^I}.
\end{equation}
Second, promoter binding and unbinding occur at fixed ligand occupancy,
\begin{equation}
A_sF
\underset{k_{-}^A}
{\overset{k_{+}^A}{\rightleftharpoons}}
A_sB,
\qquad
I_sF
\underset{k_{-}^I}
{\overset{k_{+}^I}{\rightleftharpoons}}
I_sB,
\end{equation}
with equilibrium ratio
\begin{equation}
B_A=\frac{k_{+}^A}{k_{-}^A}=B_A,
\qquad
\frac{k_{+}^I}{k_{-}^I}=B_I.
\end{equation}
Third, active-inactive conformational switching connects states with the same
ligand occupancy,
\begin{equation}
A_s b
\underset{k_{IA}^{s b}}
{\overset{k_{AI}^{s b}}{\rightleftharpoons}}
I_s b.
\end{equation}
The switching rates follow the
required equilibrium ratio,
\begin{equation}
k_{AI}^{s b}=\sqrt{\rho_{s b}},
\qquad
k_{IA}^{s b}=1/\sqrt{\rho_{s b}},
\end{equation}
where
\begin{equation}
\rho_{sF}=e^{-\Delta\epsilon_{AI}}\left(\frac{K_A}{K_I}\right)^s,
\qquad
\rho_{sB}=e^{-\Delta\epsilon_{AI}}
\left(\frac{K_A}{K_I}\right)^s\frac{B_I}{B_A}.
\end{equation}

The thermodynamic and kinetic parameters are summarized in
Table~\ref{tab:lac_mwc12_parameters}. The MWC parameters
\(K_A\), \(K_I\), \(\Delta\epsilon_{AI}\), \(R\), \(N_{NS}\), and
\(\Delta\epsilon_{RA}\) are taken from the O2, \(R=260\) LacI induction
parameterization of Ref.~\cite{razo2018tuning}. The kinetic prefactors are
not fixed by the equilibrium MWC model. We choose them so as to preserve the
MWC affinities and detailed balance while assigning an explicit relaxation
timescale to ligand binding, DNA binding, and allosteric switching. In the
calculations reported here the IPTG on-rates are
\(k_\mathrm{on}^A=k_\mathrm{on}^I=10^5\,\mathrm{M}^{-1}\mathrm{s}^{-1}\) \cite{xu2009flexibility}; the corresponding
off-rates are set by \(k_\mathrm{off}^\chi=K_\chi k_\mathrm{on}^\chi\).

\begin{table}[h]
\centering
\begin{tabular}{lll}
\toprule
Parameter & Value & Units \\
\midrule
\(K_A\) & \(139\times10^{-6}\) & \(\mathrm{M}\) \\
\(K_I\) & \(0.53\times10^{-6}\) & \(\mathrm{M}\) \\
\(\Delta\epsilon_{AI}\) & \(4.5\) & \(k_BT\) \\
\(R\) & \(260\) & dimensionless \\
\(N_{NS}\) & \(4.6\times10^{6}\) & dimensionless \\
\(\Delta\epsilon_{RA}\) & \(-13.9\) & \(k_BT\) \\
\(B_A=(R/N_{NS})e^{-\Delta\epsilon_{RA}}\) & \(61.5\) & dimensionless \\
\(B_I\) & \(10^{-3}\) & dimensionless \\
\(k_\mathrm{on}^A\) & \(10^{5}\) & \(\mathrm{M}^{-1}\mathrm{s}^{-1}\) \\
\(k_\mathrm{off}^A\) & \(13.9\) & \(\mathrm{s}^{-1}\) \\
\(k_\mathrm{on}^I\) & \(10^{5}\) & \(\mathrm{M}^{-1}\mathrm{s}^{-1}\) \\
\(k_\mathrm{off}^I\) & \(5.3\times10^{-2}\) & \(\mathrm{s}^{-1}\) \\
\(k_{+}^A\) & \(6.15\times10^{-2}\) & \(\mathrm{s}^{-1}\) \\
\(k_{-}^A\) & \(10^{-3}\) & \(\mathrm{s}^{-1}\) \\
\(k_{+}^I\) & \(10^{-3}\) & \(\mathrm{s}^{-1}\) \\
\(k_{-}^I\) & \(1.0\) & \(\mathrm{s}^{-1}\) \\
\bottomrule
\end{tabular}
\caption{Thermodynamic and kinetic parameters for the 12-state Lac
repressor MWC induction model. The equilibrium parameters reproduce the
MWC induction curve for the O2, \(R=260\) LacI simple-repression construct.
The kinetic prefactors specify one detailed-balanced CTMC realization of that
equilibrium model.}
\label{tab:lac_mwc12_parameters}
\end{table}

\subsection{Quorum Sensing}
Marine bacteria such as \textit{Vibrio harveyi} process information by secreting chemicals known as autoinducers, which signal cell density in a phenomenon known as quorum sensing \cite{mehta2009information}. A minimal model of quorum sensing incorporates three autoinducer channels, AI-1/LuxN, AI-2/LuxPQ, and CAI-1/CqsS, whose receptor signaling regulate LuxU/LuxO and ultimately the control of LuxR-dependent luminescence. The model is not intended as a fine grained representation of the quorum-sensing pathway, but it allows us to test the thermodynamic bounds on a system with multiple inputs, which are the respective autoinducer concentrations.
Each receptor possesses three states,
\begin{equation}
K_i
\underset{k_{{\rm off},i}}
{\overset{k_{{\rm on},i}\lambda_i(t)}{\rightleftharpoons}}
B_i
\underset{\nu^{-1}_i}
{\overset{\nu_i}{\rightleftharpoons}}
P_i 
\end{equation}
where \(K_i\) is a kinase-like low-cell-density state, \(B_i\) is a ligand-bound intermediate, and \(P_i\) is a phosphatase-like high-cell-density state. For a fixed autoinducer concentration $\lambda_i(t)$, the frozen equilibrium distribution is defined by the equilibrium constants between the transitions
\begin{equation}
    \pi_{K_i} = 1, \qquad 
    \pi_{B_i} = \frac{k_{{\rm off},i} \lambda_i}{k_{{\rm off},i}}, \qquad
    \pi_{P_i} = \frac{k_{{\rm off},i} \lambda_i}{k_{{\rm off},i}} \frac{\nu_i}{\nu_i^{-1}}
\end{equation}
and normalized so that $\sum_i \pi_i = 1$.

The phosphatase-state populations are then combined into a scalar signal,
\[
S(t)=\sum_i w_i p_{P_i}(t),
\]
which is mapped through a Hill function to a normalized luminescence response, according to
\begin{equation}
\hat{r}(t)= \frac{S(t)^3}{1+S(t)^3}.
\end{equation}
where a Hill coefficient of 3 was used as a phenomenological choice to produce a switch-like luminescence response.

The main assumptions of this model are receptor independence and instantaneous detailed balance of the receptor kinetics. Each autoinducer concentration \(\lambda_i(t)\) modulates only the ligand-binding transition \(K_i\to B_i\), while unbinding and conformational switching rates are fixed. Because each receptor module is a linear three-state chain with reverse transitions on every edge, the frozen generator is reversible for any fixed input value. The three channels are assumed to evolve independently until their outputs are combined at the readout level. Thus downstream phosphorelay dynamics, small-RNA regulation, LuxR accumulation, transcription, translation, and degradation are collapsed into the phenomenological Hill response rather than represented mechanistically.

The parameter values in the quorum-sensing model should be interpreted as an illustrative, scale-separated parameterization rather than a fit to measured \emph{V. harveyi} kinetic data. We choose different parameters for each receptor within representative ranges. The autoinducer association rates \(k_{\mathrm{on},i}\sim 10^5-10^6\,\mathrm{M}^{-1}\mathrm{s}^{-1}\) were chosen to lie in a plausible biochemical range for ligand-receptor binding in a crowded cellular environment \cite{xu2009flexibility}, while the dissociation rates \(k_{\mathrm{off},i}\sim 10^{-3}-10^{-2}\,\mathrm{s}^{-1}\) set effective affinities in the nM to sub-micromolar range through \(K_{D,i}=k_{\mathrm{off},i}/k_{\mathrm{on},i}\) \cite{mccready2019structural}. The ligand-bound-kinase to phosphatase conformational rates \(\nu_i,\nu_i^{-1}\sim 10^{-2}-10^{-1}\,\mathrm{s}^{-1}\) assign receptor switching times of order tens of seconds to minutes, slower than autoinducer binding, faster than the hundreds-of-minutes timescales for population growth, over which autoinducer concentrations change appreciably. 

The input scales were chosen to produce biologically reasonable autoinducer concentrations and smooth changes over a population-growth timescale. In terms of population density, the autoinducer concentrations obey
\[
\lambda_i(t)=\lambda_{i,\max}\frac{N(t)}{K_{N,i}+N(t)},
\]
where
\begin{equation}
N(t)=\frac{N_{\max}}{1+\exp[-\gamma(t-t_0)]}
\end{equation}
is the cell density that describes logistic bacterial population growth. The density scales \(K_{N,i}\sim 10^8-5\times 10^8\,\mathrm{cells/mL}\) place the response in the typical high-density regime of bacterial quorum sensing \cite{goo2024control}. In the random-protocol studies, these concentration scales set the pulse amplitudes, with pulse widths and switching times chosen over tens to hundreds of minutes so that the receptor model experiences smooth time-dependent driving. Thus the parameterization was designed to create a finite-state reversible Markov model with biologically interpretable time and concentration scales, while avoiding a claim of quantitative calibration to a specific experimental dataset.


The parameters for the model and input signals are summarized in Table \ref{tab:parameters_quorum}.

\begin{table}[h]
\centering
\begin{tabular}{lllll}
\toprule
Parameter & AI-1/LuxN & AI-2/LuxPQ & CAI-1/CqsS & Dimensions \\

$k_{\mathrm{on}}$
& $10^6$
& $10^5$
& $5\times10^5$
& $\mathrm{M^{-1}\,s^{-1}}$ \\

$k_{\mathrm{off}}$
& $10^{-2}$
& $10^{-2}$
& $5\times10^{-3}$
& $\mathrm{s^{-1}}$ \\

$\nu$
& $10^{-1}$
& $8\times10^{-2}$
& $1.5\times10^{-1}$
& $\mathrm{s^{-1}}$ \\

$\nu^{-1}$
& $10^{-2}$
& $2\times10^{-2}$
& $10^{-2}$
& $\mathrm{s}^{-1}$ \\

$w$ & 1.0 & 0.8 & 1.2 \\

$\lambda_{i,\max}$
& $100$
& $500$
& $200$
& $\mathrm{nM}$ \\

$K_{N,i}$
& $2\times10^8$
& $5\times10^8$
& $10^8$
& $\mathrm{cells\,mL^{-1}}$ \\
\bottomrule
\end{tabular}
\caption{Parameters for quorum sensing model.}
\label{tab:parameters_quorum}
\end{table}

\subsection{Nonphotochemical Quenching}
In light-harvesting photosynthetic organisms such as plants and cyanobacteria, a mechanism known as nonphotochemical quenching (NPQ) is necessary to protect the organism from oxidative damage in excess light conditions. The NPQ model based on Ref. \cite{short2022xanthophyll} is a reversible reaction network involving a series of pigments that interconvert, binding to photosynthetic proteins to quench light excitations. Unlike the previous examples, the NPQ model is not a finite-state Markov chain but rather a nonlinear reaction network possessing conservation laws.

The network involves the concentrations 
\begin{equation}
    \bm{x}=\{V,Z,P,PV,PZ,Q,\vdea \}
    \label{eq:npq-concentrations}
\end{equation}
where \(V\) is violaxanthin, \(Z\) is zeaxanthin, \(P\) is the relevant
LHCX1-like protein pool, \(PV\) and \(PZ\) are pigment-protein complexes,
\(Q\) is the active quencher, and
\[
\vdea=\frac{[\mathrm{VDE}_a]}{[\mathrm{VDE}_a]^{\mathrm{eq}}_{\mathrm{light}}}
\]
is the relative activity of violaxanthin de-epoxidase (VDE), which activates based on light intensity. We will not explicitly model the concentration of active VDE, denoted $[\mathrm{VDE}_a]$ above, but will include its activity implicitly in the $V-Z$ interconversion below. The variables are
dimensionless reduced concentrations, scaled so that the quenching-rate
constant does not appear explicitly in the output. The reaction scheme is
\[
V
\underset{k_Z}{\stackrel{k_V^\mathrm{light} \gamma}{\rightleftarrows}}
Z,
\qquad
P+V
\underset{k_{PV,b}}{\stackrel{k_{PV,f}}{\rightleftarrows}}
PV,
\qquad 
P+Z
\underset{k_{PZ,b}}{\stackrel{k_{PZ,f}}{\rightleftarrows}}
PZ,
\qquad
PZ
\underset{k_{Q,b}(\lambda)}{\stackrel{k_{Q,f}(\lambda)}{\rightleftarrows}}
Q,
\]
with a separate relaxation equation for the VDE activity derived from first-order activation-activation kinetics,
\[
\frac{d\vdea}{dt}=k_{\mathrm{VDE}}(\lambda)
\left[\vdea_{\mathrm{eq}}(\lambda)-\vdea\right],
\]
where $\gamma_\mathrm{eq}(\lambda)$ denotes the equilibrium activity in light or dark conditions. We use the relative activity as a dimensionless proxy for the concentration of VDE in the active state.

The input is a light protocol,
\begin{equation}
\lambda(t)=I(t),
\end{equation}
where we alternate piecewise light and dark sequences with variable durations and use two parameter sets associated with these conditions. 
The light input enters through the \(PZ\rightleftarrows Q\) rates and through
the equilibrium and relaxation rate of \(\vdea\). In high light, $\vdea_{\mathrm{eq}}=1$, whereas in the dark, $\vdea_{\mathrm{eq}}=k_{V,\mathrm{dark}}/k_{V,\mathrm{light}}$. Note that because $\vdea$ absorbs the light/dark dependence and is normalized by the light condition equilibrium, the rate of the $V\rightarrow Z$ reaction is given by $k_{V, \mathrm{light}} \gamma$ 

The output is the active quencher population,
\begin{equation}
\hat{r}(t)=Q(t)-Q(0).
\end{equation}
For each fixed light condition, a frozen steady state is computed and used to define the lag error. 

Parameters taken from the low light, irregular sequence model of Ref. \cite{short2022xanthophyll} are summarized in Table~\ref{tab:npq_parameters}. Initial reduced concentrations, which define the total populations, are
\[
V_0=27.0,\quad Z_0=0.197,\quad P_0=8.46\times 10^{-3}, 
\]
\[
PV_0=3.85,\quad PZ_0=2.88\times 10^{-2}, \quad Q_0=8.40\times 10^{-3}.
\]

\begin{table}[h]
\centering
\begin{tabular}{lll}
\toprule
Parameter & Value & Units \\
\midrule
\(k_{PZ,f}\) & \(6.28\) & \(\mathrm{min}^{-1}\) \\
\(k_{PZ,b}\) & \(0.364\) & \(\mathrm{min}^{-1}\) \\
\(k_{PV,f}\) & \(101\) & \(\mathrm{min}^{-1}\) \\
\(k_{PV,b}\) & \(6.00\) & \(\mathrm{min}^{-1}\) \\
\(k_{Q,f}^{\mathrm{light}}\) & \(10.9\) & \(\mathrm{min}^{-1}\) \\
\(k_{Q,b}^{\mathrm{light}}\) & \(1.51\times 10^{-2}\) & \(\mathrm{min}^{-1}\) \\
\(k_{Q,f}^{\mathrm{dark}}\) & \(1.13\) & \(\mathrm{min}^{-1}\) \\
\(k_{Q,b}^{\mathrm{dark}}\) & \(3.88\) & \(\mathrm{min}^{-1}\) \\
\(k_{V,\mathrm{light}}\) & \(0.158\) & \(\mathrm{min}^{-1}\) \\
\(k_{V,\mathrm{dark}}\) & \(4.53\times 10^{-4}\) & \(\mathrm{min}^{-1}\) \\
\(k_Z\) & \(6.21\times 10^{-2}\) & \(\mathrm{min}^{-1}\) \\
\(k_{\mathrm{VDE}}^{\mathrm{light}}\) & \(1.84\) & \(\mathrm{min}^{-1}\) \\
\(k_{\mathrm{VDE}}^{\mathrm{dark}}\) & \(1.42\) & \(\mathrm{min}^{-1}\) \\
\bottomrule
\end{tabular}
\caption{Parameter values used in the random NPQ bound studies. All rates are given in
\(\mathrm{min}^{-1}\) in the reduced model. Second-order association parameters are expressed in the corresponding reduced concentration units.}
\label{tab:npq_parameters}
\end{table}

\subsection{Random trials}
All four studies used 1000 independently generated input protocols.

The 
ion channel, Lac repressor, and quorum-sensing protocols are smooth pulse trains. Each
pulse has the form
\[
  \lambda_i(t)
  = A\,
    \frac{1}{1+\exp[-(t-t_{\mathrm{on}})/\tau]}\,
    \frac{1}{1+\exp[(t-t_{\mathrm{off}})/\tau]},
\]
with amplitude $A$, center $c$ drawn from a specified time range, and width $w$ such that
\begin{equation}
    t_\mathrm{on} = c - \frac{w}{2} \qquad t_\mathrm{off} = c + \frac{w}{2},
\end{equation}
to ensure that each pulse switches on and off smoothly rather than discontinuously.
Pulses are added together to form the overall input protocol, 
\begin{equation}
    \lambda(t) = \sum_i \lambda_i(t).
\end{equation}

\begin{table*}
\caption{Random-study parameters used by the data files entering the overlay.}
\label{tab:random_study_parameters}
\begin{ruledtabular}
\begin{tabular}{lllll}
Model & Protocol time & Points & Seed & Input family \\
NPQ low-light & \SI{20}{\minute} & BDF, $\Delta t=\SI{0.0167}{\minute}$ & 123 & alternating light/dark segments \\
Lac repressor & \SI{2000}{\second} & 1200 & 4 & smooth inducer pulses  \\
Ligand channel & \SI{50}{\milli\second} & 1000 & 4 & smooth ligand pulses  \\
Quorum sensing & \SI{800}{\minute} & 1200 & 4 & three independent autoinducer pulses  \\
\end{tabular}
\end{ruledtabular}
\end{table*}

The ion channel input is a ligand concentration over a short
\(\SI{50}{ms}\) window. Each trace begins with a baseline
\(\SI{1}{\micro M}\) and receives a uniformly random number of pulses, \(N_p\in\{1,\ldots,5\}\). The
pulse parameters are
\[
  A \sim \mathrm{Unif}(0.05,1)\times \SI{120}{\micro M}, \qquad
  w \sim \mathrm{Unif}(\SI{1.5}{ms},\SI{8}{ms}), \qquad
  \tau \sim \mathrm{Unif}(\SI{0.3}{ms},\SI{2}{ms}),
\]
with centers drawn uniformly from the central 80\% of the time window. A weak sinusoidal background is added:
\[
  A_{\mathrm{bg}}\left[\frac{1+\sin(2\pi t/T+\phi)}{2}\right],
\]
with
\[
  A_{\mathrm{bg}}\sim\mathrm{Unif}(0,\SI{20}{\micro M}), \qquad
  f\sim\mathrm{Unif}(\SI{20}{Hz},\SI{120}{Hz}), \qquad
  \phi\sim\mathrm{Unif}(0,2\pi).
\]
Concentrations are floored at \(\SI{1e-12}{M}\). Because pulses can overlap, maxima can exceed the
single-pulse amplitude scale. The generated ensemble has median maximum ligand concentration
\(\SI{88.4}{\micro M}\), with 5--95\% range
\(\SI{29.5}{\micro M}\)--\(\SI{168}{\micro M}\). The mean ligand
concentration has median \(\SI{23.5}{\micro M}\), with 5--95\% range
\(\SI{8.26}{\micro M}\)--\(\SI{44.6}{\micro M}\).

The Lac repressor input is an inducer concentration \(\lambda(t)\), in molar units, over \(\SI{500}{s}\). Each trace contains 
\(N_p \in \{1,\ldots,5\}\) smooth pulses. For each pulse,
\[
  A \sim \mathrm{Unif}(0.05,1)\times \SI{40}{\micro M}, \qquad
  w \sim \mathrm{Unif}(\SI{20}{s},\SI{180}{s}), \qquad
  \tau \sim \mathrm{Unif}(\SI{2}{s},\SI{25}{s}),
\]
and the pulse center is drawn uniformly from the central 84\% of the time window.  A
sinusoidal background is also added,
\[
  A_{\mathrm{bg}}\left[\frac{1+\sin(2\pi t/T+\phi)}{2}\right],
\]
with \(A_{\mathrm{bg}}\sim\mathrm{Unif}(0,\SI{5}{\micro M})\),
\(T\sim\mathrm{Unif}(\SI{100}{s},\SI{500}{s})\), and
\(\phi\sim\mathrm{Unif}(0,2\pi)\). Concentrations are floored at
\(\SI{1e-15}{M}\). Across the 1000 protocols, the maximum inducer concentration has median
\(\SI{35.6}{\micro M}\) and 5--95\% range
\(\SI{10.5}{\micro M}\)--\(\SI{70.1}{\micro M}\). The mean inducer
concentration has median \(\SI{12.8}{\micro M}\), with 5--95\% range
\(\SI{3.03}{\micro M}\)--\(\SI{26.9}{\micro M}\).

The quorum-sensing input consists of three independent autoinducer
concentration traces, one for each receptor channel:
AI-1/LuxN, AI-2/LuxPQ, and CAI-1/CqsS. The total time window is
\(\SI{800}{min}\). For each channel independently, the number of smooth pulses
is drawn from \(\{1,\ldots,5\}\). Pulse amplitudes are channel-scaled:
\[
  A_i \sim \mathrm{Unif}(0.05,1.5)\,\lambda_{\max,i},
\]
with \(\lambda_{\max}=\SI{100}{nM}\), \(\SI{500}{nM}\), and \(\SI{200}{nM}\) for the
three channels respectively. Pulse widths are drawn uniformly from
\(\SI{30}{min}\) to \(\SI{300}{min}\), switching times from
\(\tau\sim\mathrm{Unif}(\SI{5}{min},\SI{50}{min})\), and centers are restricted
so the pulse support lies within the simulated window. For the ensemble used in the main text figure, the maximum autoinducer concentration
over all channels has median \(\SI{642}{nM}\), with 5--95\% range
\(\SI{240}{nM}\)--\(\SI{1314}{nM}\). The mean autoinducer concentration has
median \(\SI{121}{nM}\), with 5--95\% range
\(\SI{47.5}{nM}\)--\(\SI{228}{nM}\).

The NPQ study instead uses piecewise-constant light/dark protocols with random
segment lengths.
Each NPQ input is a random on/off light protocol with total duration
\(\SI{20}{min}\). The number of segments is drawn uniformly from
\(\{4,\ldots,10\}\). Segment durations are generated by drawing independent
Gamma random variables with shape \(1.5\) and scale \(1\), adding a minimum
duration of \(\SI{0.25}{min}\) to every segment, and rescaling the durations so
that they sum to \(\SI{20}{min}\). The light state alternates between light and
dark; the initial state is chosen randomly. Thus the protocols contain
guaranteed switching rather than long single-state traces. In the generated ensemble, the number of segments has median \(7\), with a
5--95\% range of \(4\)--\(10\). The fraction of time spent in the light state
has median \(0.505\), with a 5--95\% range of \(0.266\)--\(0.748\).

\subsection{Computation of the Thermodynamic Bounds}

Although the biochemical models considered above differ substantially in their physical interpretation and mathematical structure, the evaluation of the thermodynamic bounds follows a common procedure. 
For the finite-state Markov models (ion channel, Lac repressor, and quorum sensing), the frozen equilibrium is defined by the state weights. The full entropy production is given by Eq. \ref{eq:full-entropy}. In the near-equilibrium limit, we evaluate the quadratic approximation given by Eq. \ref{eq:quad-entropy}. The output susceptibility is computed given the eigenbasis of the Markov generator by Eq. \ref{eq:susceptibility_eigenbasis}.
For the modified log-Sobolev constant, we use Eq. \ref{eq:mlsc-estimate} except in quorum sensing, where the state space is small enough that we compute the constant variationally by Eq. \ref{eq:mlsc-variational}.

The larger continuous NPQ state space requires a slightly different treatment. 
For the NPQ model, the frozen state is defined by the stationary solution of evolution equations subject to the appropriate conservation laws.
The entropy production is computed reaction by reaction,
\begin{equation}
\sigma(t)= \sum_r \left( j_r^+-j_r^-\right)\ln\frac{j_r^+}{j_r^-}
= \sum_r J_r A_r,
\label{eq:S60}
\end{equation}
where
 forward and backward fluxes for the five channels are
\begin{align}
j_{VZ}^+ &= k_V^{\rm light}\,\vdea\,V,
&
j_{VZ}^- &= k_Z\,Z,
\\
j_{PV}^+ &= k_{PV}^+PV_{\rm free}=k_{PV}^+\,[P]\,[V],
&
j_{PV}^- &= k_{PV}^-\,PV,
\\
j_{PZ}^+ &= k_{PZ}^+\,[P]\,[Z],
&
j_{PZ}^- &= k_{PZ}^-\,PZ,
\\
j_Q^+ &= k_Q^+(I)\,PZ,
&
j_Q^- &= k_Q^-(I)\,Q,
\\
j_\vdea^+ &= k_{\rm VDE}(I)\,\vdea_{\rm eq}(I),
&
j_\vdea^- &= k_{\rm VDE}(I)\,\vdea .
\end{align}
Here the \(V\to Z\) conversion is controlled by the activity variable
\(\vdea\). The \(\vdea\) channel is treated as exchange with a
light-dependent reservoir, so that its net current is
\[
J_\vdea=j_\vdea^+-j_\vdea^-
=k_{\rm VDE}(I)\,[\vdea_{\rm eq}(I)-\vdea].
\]

Near equilibrium, we use a local quadratic approximation to the entropy production. Let $\mathbf{x}_{\mathrm{ss}}(I)$ denote the frozen steady state at fixed light condition $I$. Because the NPQ network has two conservation laws, deviations from $\mathbf{x}_{\mathrm{ss}}$ are restricted to the stoichiometric subspace. Let the columns of $\mathbf{B}$ form an orthonormal basis for this subspace and write
\begin{equation}
\delta\mathbf{x}
=
\mathbf{x}-\mathbf{x}_{\mathrm{ss}}(I)
=
\mathbf{B}\mathbf{y}.
\label{eq:npq-stoichiometric-coordinates}
\end{equation}
Here $\mathbf{y}$ denotes perturbations from steady state in the reduced concentration coordinates.

Collect the forward and backward fluxes of the $R$ reversible reaction channels into the reaction-space vectors
\begin{equation}
\mathbf{j}^{\pm}(\mathbf{y};I)
=
\begin{pmatrix}
j_1^{\pm}(\mathbf{x}_{\mathrm{ss}}+\mathbf{B}\mathbf{y};I)\
\vdots\
j_R^{\pm}(\mathbf{x}_{\mathrm{ss}}+\mathbf{B}\mathbf{y};I)
\end{pmatrix},
\label{eq:npq-reaction-flux-vectors}
\end{equation}
and define the net reaction-current vector
\begin{equation}
\mathbf{J}(\mathbf{y};I)
\equiv
\mathbf{j}^{+}(\mathbf{y};I)
-
\mathbf{j}^{-}(\mathbf{y};I).
\label{eq:npq-net-reaction-current}
\end{equation}
At the frozen steady state, detailed balance implies $\mathbf{J}(\mathbf{0};I)=\mathbf{0}$. The linearized reaction currents are therefore
\begin{equation}
\mathbf{J}(\mathbf{y};I)
=
\mathbf{F}(I)\mathbf{y}
+
O(|\mathbf{y}|^2),
\label{eq:npq-linearized-reaction-current}
\end{equation}
where
\begin{equation}
\mathbf{F}(I)
\equiv
\left.
\frac{\partial\mathbf{J}}{\partial\mathbf{y}}
\right|_{\mathbf{y}=\mathbf{0}}
=
\left.
\frac{\partial(\mathbf{j}^{+}-\mathbf{j}^{-})}
{\partial\mathbf{x}}
\right|_{\mathbf{x}=\mathbf{x}_{\mathrm{ss}}}
\mathbf{B}.
\label{eq:npq-reaction-current-jacobian}
\end{equation}
Thus $\mathbf{F}\in\mathbb{R}^{R\times d}$ maps a displacement in the $d$-dimensional stoichiometric subspace to the corresponding $R$ reaction currents.

At the frozen steady state, detailed balance further implies
\begin{equation}
j_{r,\mathrm{ss}}^{+}
=
j_{r,\mathrm{ss}}^{-}
\equiv
j_{r,\mathrm{ss}}
\qquad
\text{for each reaction } r.
\label{eq:npq-detailed-balance-fluxes}
\end{equation}
Define the diagonal matrix in reaction space
\begin{equation}
\mathbf{D}(I)
=
\operatorname{diag}
\left(
\frac{1}{j_{1,\mathrm{ss}}},
\ldots,
\frac{1}{j_{R,\mathrm{ss}}}
\right).
\label{eq:npq-reaction-flux-metric}
\end{equation}
The reaction affinities $\mathbf{A}$ then have the linear expansion
\begin{equation}
\mathbf{A} = \ln\left(
\frac{\mathbf{j}^{+}}{\mathbf{j}^{-}}
\right)
=
\mathbf{D}(I)\mathbf{F}(I)\mathbf{y}
+
O(|\mathbf{y}|^2),
\label{eq:npq-linearized-affinity}
\end{equation}
where the logarithm and ratio are understood component-wise.

Substituting the linearized currents and affinities into the reaction-wise entropy production gives
\begin{equation}
\sigma
=
\mathbf{y}^{T}
\mathbf{F}^{T}\mathbf{D}\mathbf{F}
\mathbf{y}
+
O(|\mathbf{y}|^3).
\label{eq:npq-quadratic-entropy-production}
\end{equation}
We therefore define the quadratic entropy-production matrix in the stoichiometric subspace as
\begin{equation}
\mathbf{G}(I)
\equiv
\mathbf{F}(I)^{T}\mathbf{D}(I)\mathbf{F}(I),
\qquad
\sigma
=
\mathbf{y}^{T}\mathbf{G}(I)\mathbf{y}
+
O(|\mathbf{y}|^3).
\label{eq:npq-entropy-production-matrix}
\end{equation}

The lag in the NPQ output is the deviation in the active quencher concentration,
\begin{equation}
\epsilon_\mathrm{lag} = \delta Q
=
\mathbf{c}^{T}\delta\mathbf{x}
=
\mathbf{c}^{T}\mathbf{B}\mathbf{y}
=
\mathbf{b}^{T}\mathbf{y},
\qquad
\mathbf{b}
\equiv
\mathbf{B}^{T}\mathbf{c},
\label{eq:npq-output-lag}
\end{equation}
where $\mathbf{c}=\mathbf{e}_{Q}$ selects the $Q$ coordinate in concentration space.

Assuming $\mathbf{G}$ is positive definite on the stoichiometric subspace, we may write
\begin{equation}
\delta Q
=
\left(\mathbf{G}^{-1/2}\mathbf{b}\right)^{T}
\left(\mathbf{G}^{1/2}\mathbf{y}\right).
\label{eq:npq-output-metric-factorization}
\end{equation}
Applying the Cauchy--Schwarz inequality gives
\begin{equation}
(\delta Q)^2
\leq
\left(
\mathbf{b}^{T}\mathbf{G}^{-1}\mathbf{b}
\right)
\left(
\mathbf{y}^{T}\mathbf{G}\mathbf{y}
\right)
=
\tau_m \sigma
+
O(|\mathbf{y}|^3).
\label{eq:npq-quadratic-bound}
\end{equation}
The NPQ susceptibility entering the quadratic bound is therefore
\begin{equation}
\tau_m(I)
=
\mathbf{b}^{T}\mathbf{G}^{-1}\mathbf{b}
=
\left(\mathbf{B}^{T}\mathbf{c}\right)^{T}
\mathbf{G}^{-1}
\left(\mathbf{B}^{T}\mathbf{c}\right).
\label{eq:npq-susceptibility}
\end{equation}

Because we have restricted $\mathbf{G}$ to the stoichiometric subspace, it has no zero eigenvalue and is invertible.

\begin{figure}
    \centering
    \includegraphics[width=1\linewidth]{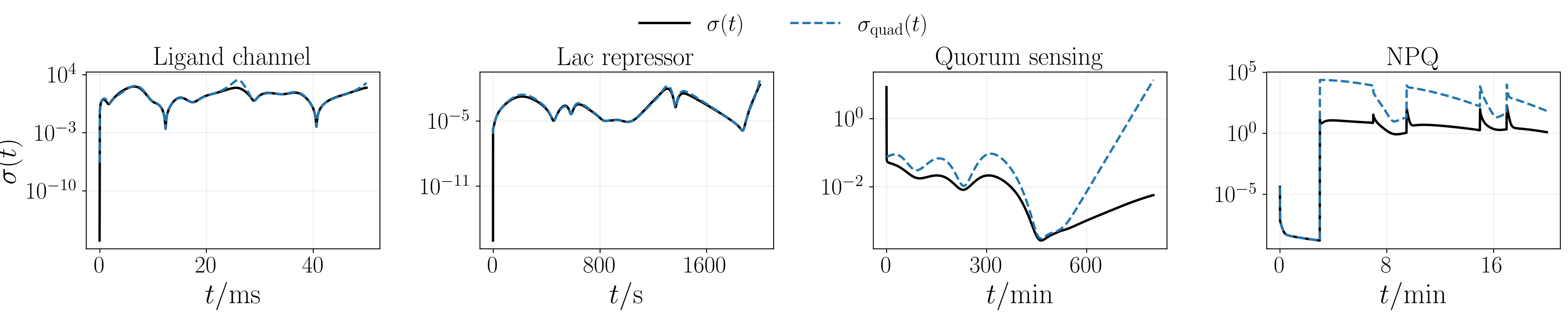}
\caption{
Comparison between the near equilibrium and full calculation of the entropy production for the four reversible biological Markov models. }
\label{fig:bio-entropy-compare}
\end{figure}

The quadratic approximation to the entropy production for the four biological Markov models is compared to its full evaluation in Fig.~\ref{fig:bio-entropy-compare}. For the Lac repressor and Ligand channel models, the quadratic approximation is accurate across the timeseries. This is less so for the quorum sensing and NPQ models. In both of the latter models, the qualitative behavior is consistent between the two estimators, but the quadratic approximation systemically over-estimates the full entropy production. In the quorum sensing model this is due to the system exploring configurations with low probability states, for which the ratio $p_i/\pi_i$ becomes large. In the NPQ model, this is due to the abrupt change in the input being discontinuous, and outside the strict adiabatic approximation. Smoothing the light intensity changes over 1 min does not significantly reduce the difference, as the NPQ model has a long memory due to metastability in the Markov network. 

Finally, we define a local proxy for the modified log-Sobolev constant using the corresponding inequality. The Sobolev constant is defined variationally by Eq. \ref{eq:mlsc-variational}, but computing it numerically was intractable for NPQ. Instead, we use the quadratic proxies for the relative and total entropy that satisfy Eq. \ref{eq:S10}. For concentrations close to the frozen steady state $\mathbf{x}_{\mathrm{ss}}$, the relative entropy has the expansion
\begin{equation}
D(\mathbf{x}|\mathbf{x}_{\mathrm{ss}})
=
\frac{1}{2}
\delta\mathbf{x}^{T}
\mathbf{H}
\delta\mathbf{x}
+
O(|\delta\mathbf{x}|^3),
\qquad
\mathbf{H}
=
\operatorname{diag}
\left(
\frac{1}{x_{1,\mathrm{ss}}},
\ldots,
\frac{1}{x_{N,\mathrm{ss}}}
\right),
\label{eq:npq-relative-entropy-hessian}
\end{equation}
where $\mathbf{H}$ is the Hessian of the relative entropy evaluated at $\mathbf{x}_{\mathrm{ss}}$.
In the stoichiometric tangent space, the relative entropy is
\begin{equation}
D(\mathbf{x}|\mathbf{x}_{\mathrm{ss}})
=
\frac{1}{2}
\mathbf{y}^{T}
\mathbf{H}_{s}
\mathbf{y}
+
O(|\mathbf{y}|^3),
\qquad
\mathbf{H}_{s}
\equiv
\mathbf{B}^{T}\mathbf{H}\mathbf{B}.
\label{eq:npq-reduced-entropy-hessian}
\end{equation}
In the local quadratic approximation, the modified log-Sobolev inequality (Eq. \ref{eq:S10}) gives
\begin{equation}
\alpha_{\mathrm{loc}}
\leq
\frac{
\mathbf{y}^{T}\mathbf{G}\mathbf{y}
}{
\mathbf{y}^{T}\mathbf{H}_{s}\mathbf{y}
}.
\label{eq:npq-local-sobolev-rayleigh}
\end{equation}
The largest constant for which the local inequality holds is therefore the minimum generalized eigenvalue
\begin{equation}
\alpha_{\mathrm{loc}}
=
\lambda_{\min}
\left(
\mathbf{G}_{s},
\mathbf{H}_{s}
\right)
=
\lambda_{\min}
\left(
\mathbf{H}_{s}^{-1/2}
\mathbf{G}_{s}
\mathbf{H}_{s}^{-1/2}
\right).
\label{eq:npq-local-sobolev-proxy}
\end{equation}

This quantity measures the slowest local rate of entropy dissipation relative to the relative-entropy curvature over perturbations compatible with the conservation laws. It should be interpreted as a near-steady-state analog of the modified log-Sobolev constant, rather than as a rigorous global modified log-Sobolev constant for the nonlinear NPQ dynamics.

\begin{figure}
    \centering
    \includegraphics[width=.5\linewidth]{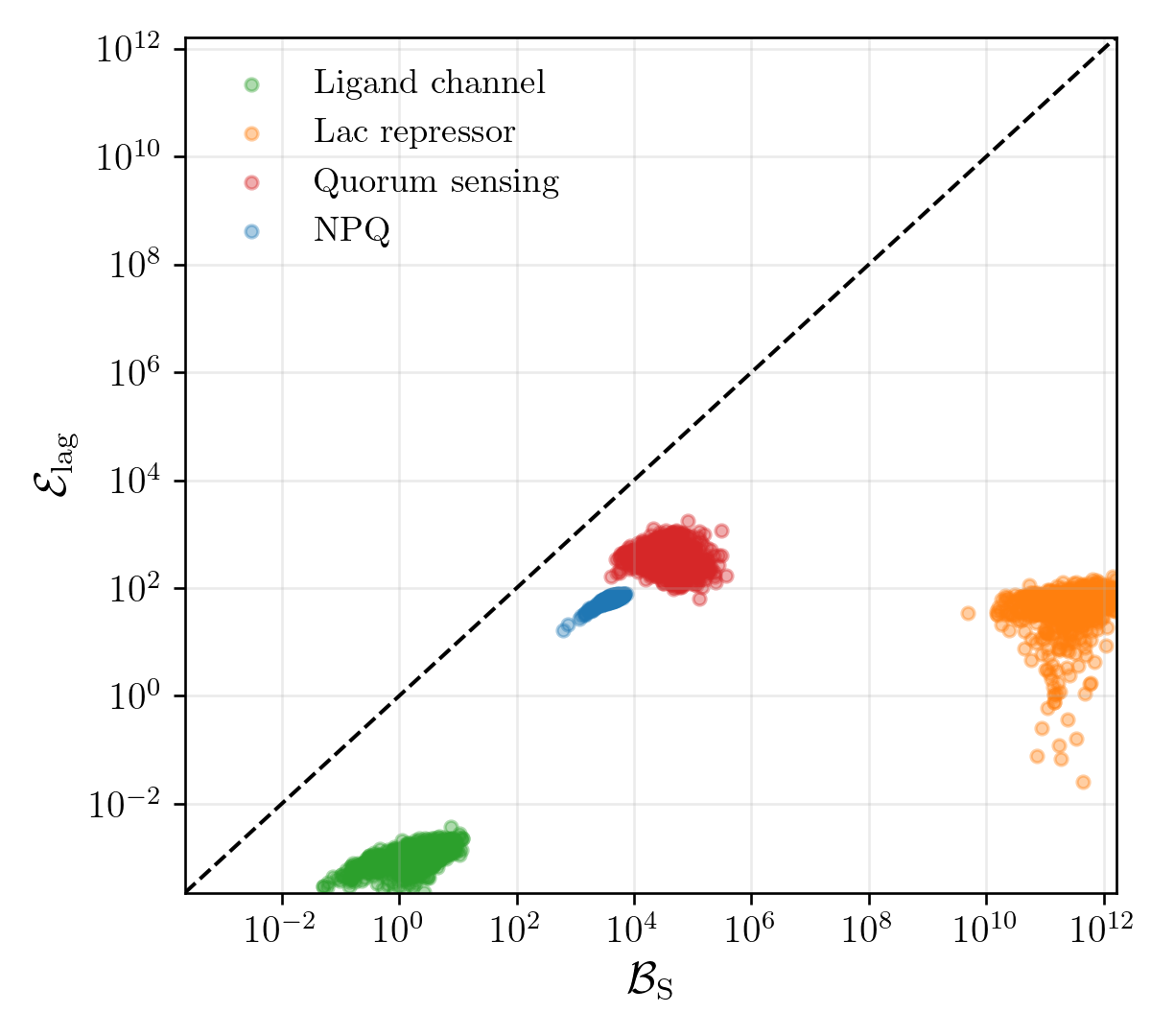}
\caption{
Comparison of the integrated lag error with the integrated Sobolev bound across 1000 random input protocols for each model. Across models and input ensembles, the lag error
remains below the thermodynamic bound.
}
\label{fig:bio-sobolev}
\end{figure}

The main text demonstrates that the near equilibrium bound provides a tight constraint on the integrated lag error, $\mathcal{E}$. In Fig.~\ref{fig:bio-sobolev} we compare the integrated Sobolev thermodynamic bound,
\begin{equation}
\mathcal B_\mathrm{S} =\int dt \,\frac{\sigma(t)}{4 \mlsc(t)}
\end{equation}
to the lag error. We find that the bound is satisfied for all models, across all random trials, but much looser than the near equilibrium result. 

\section{Numerical Methods}

To propagate the master equation dynamics for each system, we used explicit matrix exponentiation where tractable (10-state Markov chain, quorum sensing). For the ligand channel and Lac repressor we use an explicit 4th order Runge-Kutta method. NPQ requires a stiff ODE solver to handle nonlinear transitions, so we use the backward differentiation formula.
At each integration step, the probability vector was renormalized
to eliminate the accumulation of roundoff errors. 
The time integrals were evaluated using the trapezoidal rule.

\bibliography{ref}